\documentclass[manuscript]{acmart}

\AtBeginDocument{%
  \providecommand\BibTeX{{%
    \normalfont B\kern-0.5em{\scshape i\kern-0.25em b}\kern-0.8em\TeX}}}

\setcopyright{acmcopyright}
\copyrightyear{2021}
\acmYear{2021}
\acmDOI{}

\acmConference[WebSci '21]{13th ACM Web Science Conference 2021}{June 21--25, 2021}{Southampton, UK}
\acmBooktitle{13th ACM Web Science Conference 2021,
  June 21--25, 2021, Southampton, UK}
\acmISBN{}

\usepackage{paralist}
\usepackage{subfig}

\begin{document}

\title{Limiting Tags Fosters Efficiency}
%Your title must be in mixed case, not sentence case. 
% That means all verbs (including short verbs like be, is, using,and go), 
% nouns, adverbs, adjectives should be capitalized, including both words in hyphenated terms, while
% articles, conjunctions, and prepositions are lower case unless they
% directly follow a colon or long dash

\author{Tiago Santos}
\email{tsantos@iicm.edu}
\affiliation{
    \institution{Institute of Interactive Systems and Data Science, Graz University of Technology}
}
\author{Keith Burghardt}
\email{keithab@isi.edu}
\affiliation{
    \institution{Information Sciences Institute, University of Southern California}
}
\author{Kristina Lerman}
\email{lerman@isi.edu}
\affiliation{
    \institution{Information Sciences Institute, University of Southern California}
}
\author{Denis Helic}
\email{dhelic@tugraz.at}
\affiliation{
    \institution{Institute of Interactive Systems and Data Science, Graz University of Technology}
}

\begin{abstract}
Tagging facilitates information retrieval in social media and other online communities by allowing users to organize and describe online content. Researchers found that the efficiency of tagging systems steadily decreases over time, because tags become less precise in identifying specific documents, i.e., they lose their descriptiveness. However, previous works did not answer how or even whether community managers can improve the efficiency of tags. In this work, we use information-theoretic measures to track the descriptive and retrieval efficiency of tags on Stack Overflow, a question-answering system that strictly limits the number of tags users can specify per question. We observe that tagging efficiency stabilizes over time, while tag content and descriptiveness both increase. To explain this observation, we hypothesize that limiting the number of tags fosters novelty and diversity in tag usage, two properties which are both beneficial for tagging efficiency. To provide qualitative evidence supporting our hypothesis, we present a statistical model of tagging that demonstrates how novelty and diversity lead to greater tag efficiency in the long run. Our work offers insights into policies to improve information organization and retrieval in online communities.
\end{abstract}




\maketitle
%
%--------------------------------------------------
%
\section{Introduction}
Social tagging, a popular form of content annotation, helps users in online communities categorize and retrieve information. Unlike traditional methods for organizing information that are based on predefined categories or ontologies, tagging enables people to label content with free-form terms. The knowledge organization emerging from the tagging activities of many people can better adapt to innovations and the growing complexity of knowledge than traditional methods~\cite{cattuto2008,furnas2006tagging,markines2009,Willet2011}. Research into the properties of tagging systems for information organization and retrieval has highlighted the capabilities and limitations of tags' information structuring properties~\cite{schmitz2006, markines2009, helic2011,LEY2015,Mamykina2011}, information seeking qualities~\cite{chi2008understanding,chi2008information,hotho2006,dimitrov2018tag,krestel2009latent,klavsnja2018}, and navigability~\cite{kammerer2009signpost,helic2010navigability}.
Nevertheless, optimizing tagging systems to improve the organization, retrieval and visualization of content still attracts substantial research effort~\cite{XIE2014,dimitrov2018tag,kowald2017temporal,Meo2014,trattner2016,klavsnja2018}.

Chi \& Mytkowicz made an attempt to quantify the capacity of tags to precisely identify documents---a property they called \textit{tag efficiency}~\cite{chi2008understanding}. They proposed an information theoretic measure of tag efficiency and showed that it decreases over time on the social bookmarking site del.icio.us, an early example of a social tagging system that allowed any user to add free-form labels to web pages. Their finding implies that tags became less useful over time for the purpose of information retrieval on del.icio.us. However, the study did not specify whether  this was a generic property of social tagging systems, or how system managers could counter this trend to improve tag efficiency. 

\noindent \textbf{This work.}
In this paper, we observe improved information organization and retrieval in a tagging system that imposes a limit on the number of tags used to annotate documents. 
Limiting the number of tags and at same time simplifying the creation of composite tags, may nudge users to create fewer but more specific tags, which improves tag efficiency~\cite{koerner2010,Zubiaga2011}. 
We study a large-scale dataset of tagged questions from Stack Overflow,\footnote{\url{https://stackoverflow.com/questions}} a popular online community dedicated to answering questions (Q\&A) on programming-related topics.
On Stack Overflow, users can annotate each question they ask with up to five tags. 
We use information-theoretic measures to quantify the capacity of tags to organize and retrieve information. 
Specifically, we compute tag entropy, the conditional entropy of questions given tags, and the mutual information of questions and tags as measures for information content, information retrieval, and tag descriptiveness, respectively. These three measures assess \textit{tag efficiency}. 
Finally, we explore the mechanism leading to observed trends by simulating a statistical model of tagging. The model, which is inspired by models of Web growth~\cite{kleinberg1999} and evolution of biological and technological systems~\cite{tria2013}, captures the growth of tags through novelty, diversity, and reinforcement.

\noindent \textbf{Findings and contributions.} 
While the amount and complexity of information on Stack Overflow grows over time, increasing the demands for  organizing that information, our findings indicate that capacity of tags to retrieve information  
stabilizes after an initial period of decline. 
Further, we also observe a steady \textit{increase} in tag descriptiveness. This is in contrast to an earlier study of social tagging which showed a \textit{decreasing} capacity of tags to describe and identify documents~\cite{chi2008understanding}.
We hypothesize that improvement in tag efficiency is linked to strict limits on the number of 
tags imposed by Stack Overflow. 
We show that questions with fewer than five tags exhibit even higher tag efficiency over time. 
The difference in tag efficiency is most pronounced in the comparison between questions with five tags and those with only one tag, and this difference becomes smaller when the number of tags increases. 
Limiting tags therefore appears to encourage users to (i) create new tags by concatenating existing ones, based on the implicit tag hierarchy on Stack Overflow and to (ii) diversify their tag selection by using more specific rather than popular tags. The additional effort in annotating content contributes towards higher tag efficiency while reducing the number of tags to describe a question.
Finally, with our statistical model  
we (i) qualitatively recreate a range of behaviors observed in real systems, and (ii) illustrate how novelty and diversity in tag usage improve tagging efficiency. This qualitative evidence thus supports our empirically-grounded information retrieval hypotheses. 

Our results are of practical relevance for managers aiming to optimize their tagging systems.
Specifically, our findings suggest that limiting 
the number of tags people can use to annotate content may improve the efficiency of tags for information retrieval.
However, system managers may wish to consider the trade-offs between simplifying information organization (e.g., by recommending popular tags)
and encouraging elaborate content annotation (e.g., through manual creation of descriptive tags), as we link gains in information organization to decreasing tag efficiency.

%
%--------------------------------------------------
%
\section{Related Work}

\noindent \textbf{Social Tagging.} Tagging, a way to categorize content in online platforms, has witnessed broad usage in content aggregation websites, such as del.icio.us~\cite{furnas2006tagging}, and social networks, such as Twitter or Instagram~\cite{zhang2019language}, via the hashtag mechanism. Previous work has studied the function of tags for information seeking purposes~\cite{chi2008information, Heymann2010}, as well as for social advocacy, such as \#MeToo~\cite{mendes2018metoo} and \#BlackLivesMatter~\cite{olteanu2016characterizing}. Social tagging systems vary on who can tag (document owner or anyone), what tags are used (freeform or dictionary terms), and how many tags can be added.

\noindent \textbf{Information organization.}
Furnas et al.~\cite{furnas2006tagging} see the social tagging process as a collective effort of organizing information without predefined term vocabulary. 
Several studies analyzed what motivates users to create tags in order to design better user interfaces~\cite{lin2015}, or model information retrieval. For example, K\"orner et al.~\cite{koerner2010} characterized users according to their tagging motivations
and concluded with the observation that the semantics, and hence, information retrieval efficiency is best achieved through verbose tags. 
Collaborative systems for knowledge creation (e.g., Wikipedia or Stack Overflow) benefit from shared information organization tools such as social tagging as they support establishment of common domain understanding~\cite{Willet2011}. While automatic extraction of keywords using modern natural language processing and deep learning methods are frequently applied to support information organization efforts (e.g., by presenting a set of keywords for users to select from)~\cite{Awawdeh2010, bagheri2016, MIOTTO2013}, establishment of a shared vocabulary among users supports individual and collaborative sensemaking~\cite{Chen2011, Mamykina2011, LEY2015}. In particular, Mamykina et al. showed the enhanced ability of individual users of social tagging systems to recollect relevant information, and stressed the importance of controlled vocabularies and expert moderation~\cite{Mamykina2011}, whereas Bagheri and Ensan~\cite{bagheri2016} showed the abilities of tags in organizing content in collaborative software projects.
In another line of work, Golder and Huberman~\cite{Golder2006} found that 
tags exhibit a strong popularity bias. More recently, Trattner et al.~\cite{trattner2016} 
found a recency bias, in which tag usage strongly depended on time.

In our work, we ask whether and how user interface design choices, such as limiting the number of user-assigned tags, support information organization in the long term.

\noindent \textbf{Information retrieval.} 
There are two main streams of research on improving information retrieval with tags. 
Firstly, tag recommender systems aim to predict tags for users on a document of interest. Building on seminal work of Heymann et al.~\cite{heymann2008social}, 
tag recommendation systems have evolved to incorporate models and insights from natural language processing~\cite{krestel2009latent} or cognitive science~\cite{kowald2017temporal}. 
Further, tags help in profiling users in recommender systems~\cite{zhang2011}. For instance, Klasnja-Milicevic~\cite{klavsnja2018} enhanced a standard tensor factorization method to improve recommendations in an educational setting, Enrich et al.~\cite{enrich2013} applied cross-domain tags to remedy cold-start problem in recommender systems, and Zuo et al.~\cite{zuo2016} applied auto-encoders and deep learning to extract more precise user profiles for recommendations.
Secondly, considering tag navigation, previous work proposed a tag exploration system to help users grasp certain topics~\cite{kammerer2009signpost}. Helic et al.~\cite{helic2010navigability}, however, challenged the underlying hypothesis that tagging systems also support efficient navigation. In a later work, Helic and Strohmaier proposed an algorithm to improve navigational efficiency of tag hierarchies~\cite{helic2011}.

We position our work in the second stream of research, as we inspect the efficiency of design choices on tag-based information retrieval. In particular, a study of del.icio.us, which allowed users to annotate Web pages with tags, demonstrated that the efficiency of tags for information retrieval declined, as indicated by increasing entropy (i.e., uncertainty) of a document given a set of tags used to describe it~\cite{chi2008understanding}. We extend this work with a longitudinal study of tag efficiency in a system that limits the number of tags users can apply to a document, to examine if this improves tag efficiency over time. 
Our research thus offers evidence-based design recommendations for tagging system managers. Orthogonal research~\cite{gligoric2018constraints} found a link between constraints and creative production on social media, therefore 
it may not be clear \emph{a priori} whether limiting the number of tags supports information retrieval. Does a tag limit stifle users' creativity or do users adhere to the limit and employ specific and descriptive tags? 
We find evidence for the latter.

\noindent \textbf{Improving tagging systems.}
Researches have proposed a variety of ways to improve social tagging systems.
Cantador et al.~\cite{CANTADOR2011} propose mapping of tags to a set of predefined ontology classes from external sources such as WordNet or Wikipedia with the goal of reducing noise in tag collections. Similarly, Zubiaga et al.~\cite{Zubiaga2011} suggest that clustering user tagging profiles supports categorization of tagged resources. Further, Meo et al.~\cite{Meo2014} compare tagging and social behavior of the users with the semantics emerging from the tags that they use. Their findings support applications such as recommender systems, user profile merging and estimation of user similarity across platforms. On the other hand, Xie et al.~\cite{XIE2014} extract communities of similar users from their tagging profiles to enrich future tag suggestions. Jabeen et al.~\cite{jabeen2014} provide a comprehensive review of approaches to extracting semantics from social tagging datasets.

Our work extends this line of research by analyzing how user interface design influences and potentially improve the information retrieval efficiency of tagging systems.

\noindent \textbf{Growth models for information systems.} Finally, stochastic models for information systems are critical to better understand the patterns seen in data. Kleinberg et al.~\cite{kleinberg1999} designed a model of the Web growth in which the new Web pages are created iteratively with pages stochastically deciding whether to copy the links from an already existing page or to create new links uniformly at random. This preferential attachment mechanism~\cite{Barabasi1999} results in a rich-get-richer dynamics and a skewed degree distribution. Tria et al.~\cite{tria2013} developed a model that extends a basic preferential attachment mechanism by also modeling the time correlation in novelties, such as new Wikipedia webpages.

We explore a model of the growth of resources and tags, which reproduces a broad range of  behaviors observed in real-world social tagging systems.

%
%--------------------------------------------------
%

\section{Materials and Methods}

\begin{figure}[t]
	\centering
	\subfloat[New questions per month]{\includegraphics[width=0.24\textwidth]{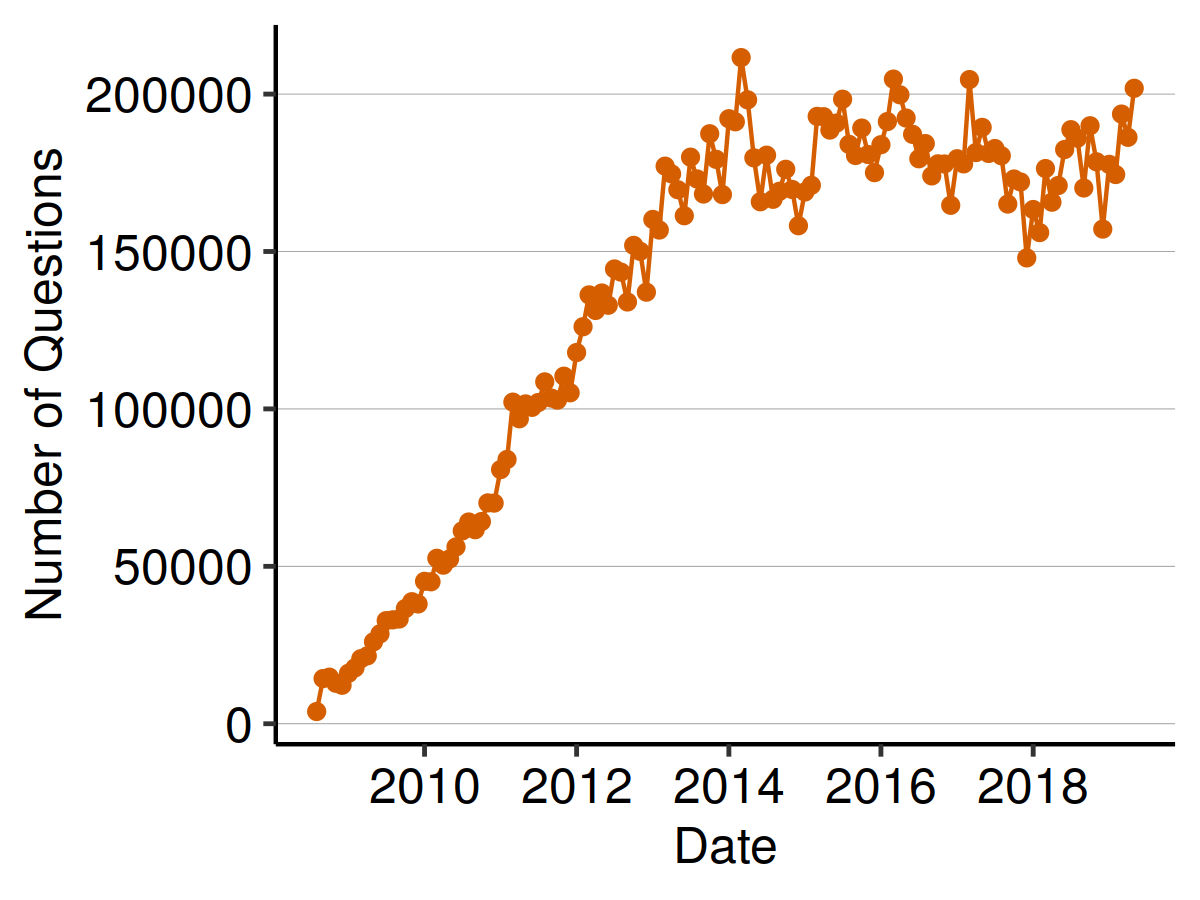}}
	\subfloat[Tags per question and month]{\includegraphics[width=0.24\textwidth]{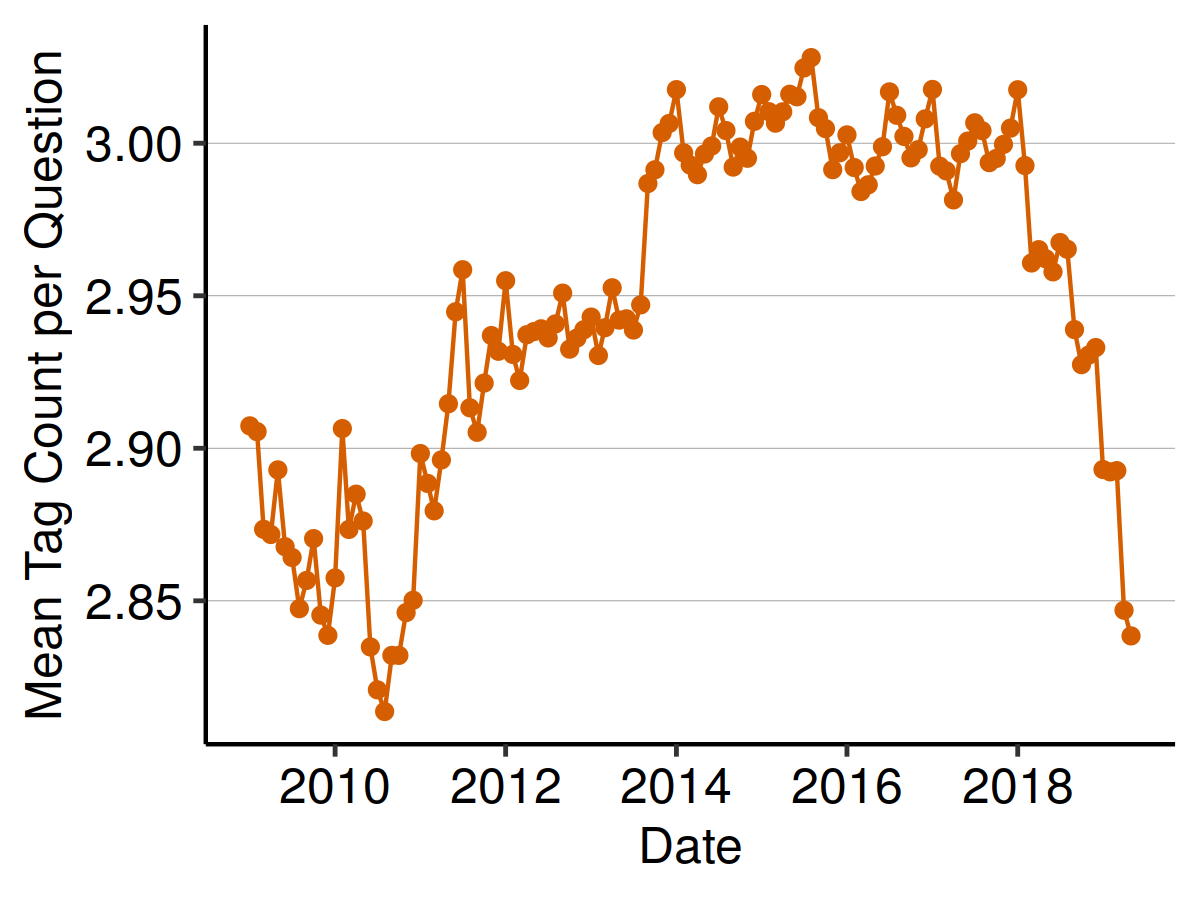}}
	\caption{\textit{Growth of questions and tags on Stack Overflow.} The number of new questions and tags increases steadily from August $2008$ until May $2019$. 
		(a) The number of new questions per month, as a proxy for the amount of information retrieved, stabilizes at  $150-200$k, following an initial phase of strong growth.
		(b) Users assign on a monthly average about three tags per question. 
		}
	\label{fig:data}
\end{figure}

\noindent \textbf{Data.} We study Stack Overflow, a large Q\&A community dedicated to answering questions related to programming. We obtained data on Stack Overflow from August 2008 through May 2019\footnote{Data source: \url{https://archive.org/download/stackexchange/}}, comprising $17.7$ million questions.
Some descriptive statistics of the growth of questions and tags in the data are shown in Figure~\ref{fig:data}.
We focus on questions and tags users employ to categorize questions. 
Users can assign up to five unique tags per question by picking terms from a pre-defined vocabulary.
Only experienced users with enough reputation on Stack Overflow\footnote{\url{https://stackoverflow.com/help/privileges/create-tags}} may propose new tags.

To analyze the tag efficiency in Stack Overflow we study the evolution of 
\begin{inparaenum}[(a)]
\item \textit{information content} (quantifying how much information is contained in the system),
\item \textit{information retrieval} (measuring the difficulty of finding a question via tags), and
\item \textit{tag descriptiveness} (quantifying the utility of tags in describing questions)
\end{inparaenum}
over time.
We use information-theoretic measures to compute these quantities, following the work by Chi and Mytkowicz~\cite{chi2008understanding}.

\noindent \textbf{Information content.}
In addition to measuring a tagging system's growth as counts over time, 
we also consider how the \textit{distributions} of tags and questions change
by computing their entropy. Mathematically, the entropy is given by:
\begin{equation} \label{eq:ent_Q}
H(Q) = - \sum_{q \in \mathcal{Q}} p(q) \log p(q),\text{ }H(T) = - \sum_{t \in \mathcal{T}} p(t) \log p(t).
\end{equation}

Entropy measures the information content of a random variable by computing the number of bits needed to encode the uncertainty in that random variable. Thus, this metric captures uncertainty in choosing a question or tag from the empirical distribution of questions or tags. For example, do users prefer to use popular or niche tags? The former case corresponds to a low level of entropy (or equivalently a low information content or a low level of uncertainty when selecting a tag at random) as many users simply reuse a popular tag over and over again. In contrast, the infrequently-used niche tags  are described by a high level of entropy. 

Question and tag entropy that \textit{grows with time} are therefore desirable properties of a tagging system as they signal that the amount and complexity of information in the system is increasing. On the other hand, constant or decaying question and tag entropy indicate a saturated system in which no new information is added. This can either happen if no new questions or tags are created (constant entropy) or if the distributions become heavily skewed with time (constant or decaying entropy) due to, for example, strong popularity bias.

In a question-answering system, the probability of randomly choosing a question is inversely proportional to the number of questions $|\mathcal{Q}|$, which usually appear only once~\cite{ahasanuzzaman2016mining}.
Therefore, $H(Q)$ corresponds to the entropy of a uniform distribution, which is $\log |\mathcal{Q}|$.
Tags, however, do not occur only once, and hence $p(t)$ is the relative frequency of a tag $t$.

\noindent \textbf{Information retrieval.}

We use the conditional entropy of questions given tags as a proxy for tag retrieval efficiency.
Intuitively, this metric captures the uncertainty in
identifying a question after a specific tag has been selected by a user. The lower this uncertainty the more efficiently tags can identify questions. As such, the conditional entropy measures how difficult is to retrieve a question given a set of tags and preferably this conditional entropy \textit{decays with time}.

Formally, we define the conditional entropy of questions $q \in \mathcal{Q}$ given tags $t \in \mathcal{T}$ as follows:
\begin{equation} \label{eq:condent}
H(Q|T) = - \sum_{t \in \mathcal{T}} p(t) \sum_{q \in \mathcal{Q}} p(q|t) \log p(q|t),
\end{equation}
where $p(t)$ is as defined above and $p(q|t)$ is the conditional probability of a question $q$ given a tag $t$, i.e., the probability of picking a single question out of all questions employing a certain tag, or, mathematically, one divided by the number of questions with a given tag $t$. The calculation of $p(q|t)$ only holds for Stack Overflow where distinct tags are added to each question by a single creator of that question. In systems where multiple users can tag resources with repeating tags $p(q|t)$ corresponds to the conditional relative frequency of a question given the tag~\cite{chi2008understanding}.

The conditional entropy is minimized at zero---this happens when tags uniquely identify questions (one-to-one mapping to questions), i.e., when the information of tags suffices to retrieve questions with no uncertainty left.

\noindent \textbf{Tag descriptiveness.}
Our third information-theoretic measure captures the descriptiveness of tags, i.e., how much tags tell us about questions.
This metric is the mutual information between questions and tags, defined as:
\begin{equation} \label{eq:mutual_info}
I(Q;T) = H(Q) - H(Q|T) 
\end{equation}
This measure is akin to a distance between the question entropy and the conditional entropy of questions given tags, i.e., it quantifies the reduction of uncertainty in questions due to the knowledge of tags. If it is large, then knowledge of tags greatly reduces uncertainty in questions, and thus we say tags are very descriptive of questions. Otherwise, tags do not convey much information about questions as they are largely independent of them. 

In an efficient tagging system mutual information \textit{grows over time}, meaning that with time tags tell us more about questions. Note that this seamlessly fits with the other two desirable properties of a tagging system:
\begin{inparaenum}[(a)]
\item growing $H(Q)$, and
\item decaying $H(Q|T)$ lead
\end{inparaenum}
automatically to an increase in $I(Q;T)$.
Mathematically, we can also write $I(Q;T) = H(T) - H(T|Q)$, an equation which we will use to analyze our empirical results in more details.

In this work, we inspect the longitudinal development of the above defined quantities on Stack Overflow. Specifically, we compute those metrics each month from August $2008$ to May $2019$ on the growing sets of questions $\mathcal{Q}$ and of tags $\mathcal{T}$.

%
%--------------------------------------------------
%

\begin{figure*}[t]
	\centering
	\subfloat[Inf. content (questions)]{\includegraphics[width=0.24\textwidth]{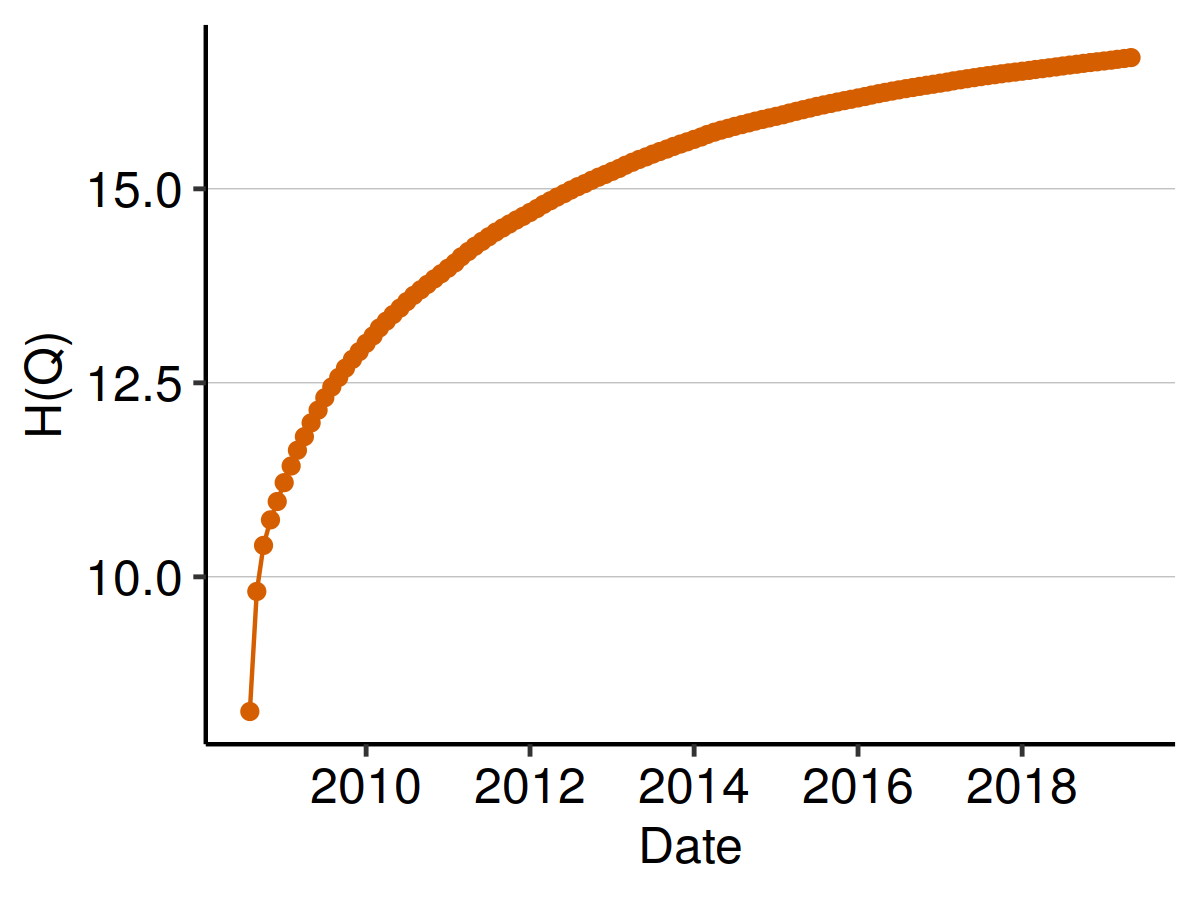}}	\subfloat[Inf. content (tags)]{\includegraphics[width=0.24\textwidth]{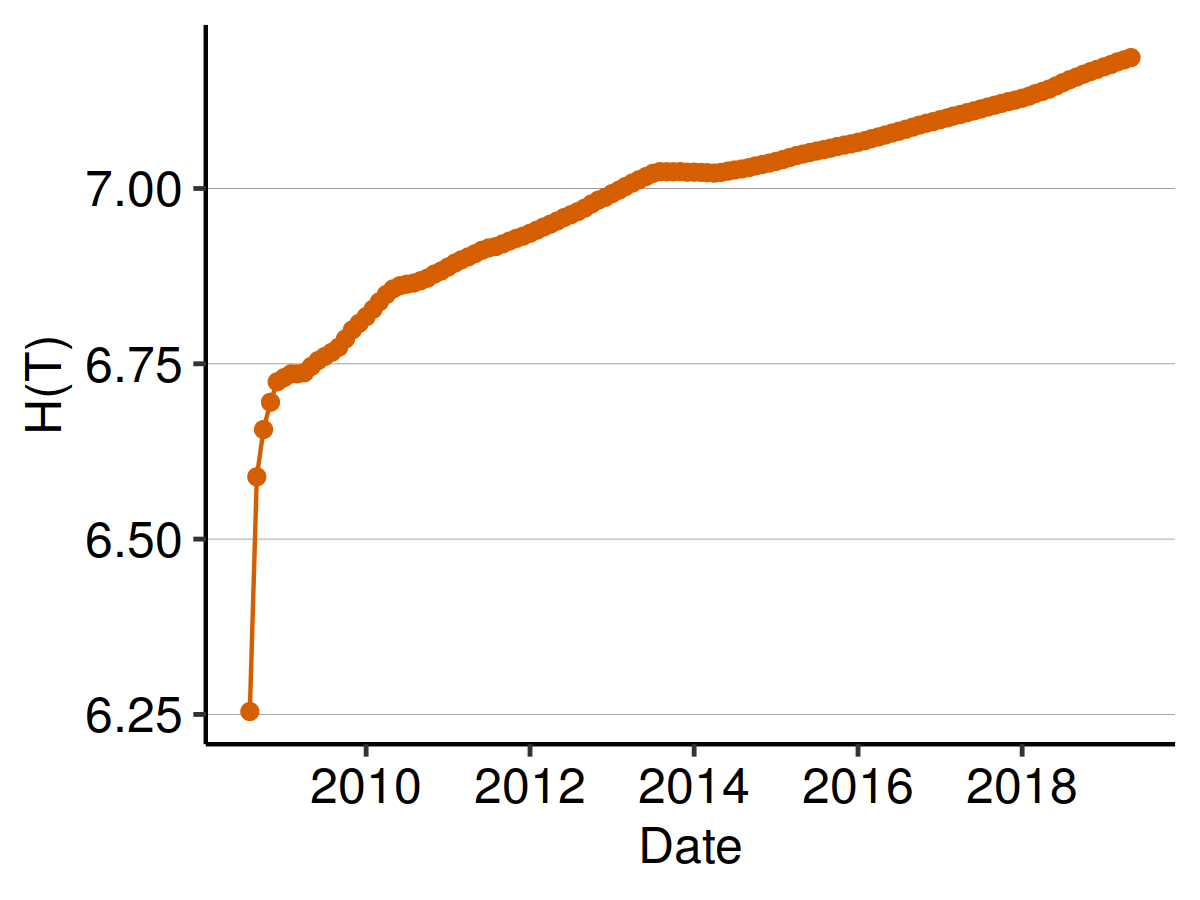}}
	\subfloat[Difficulty of inf. retrieval]{\includegraphics[width=0.24\textwidth]{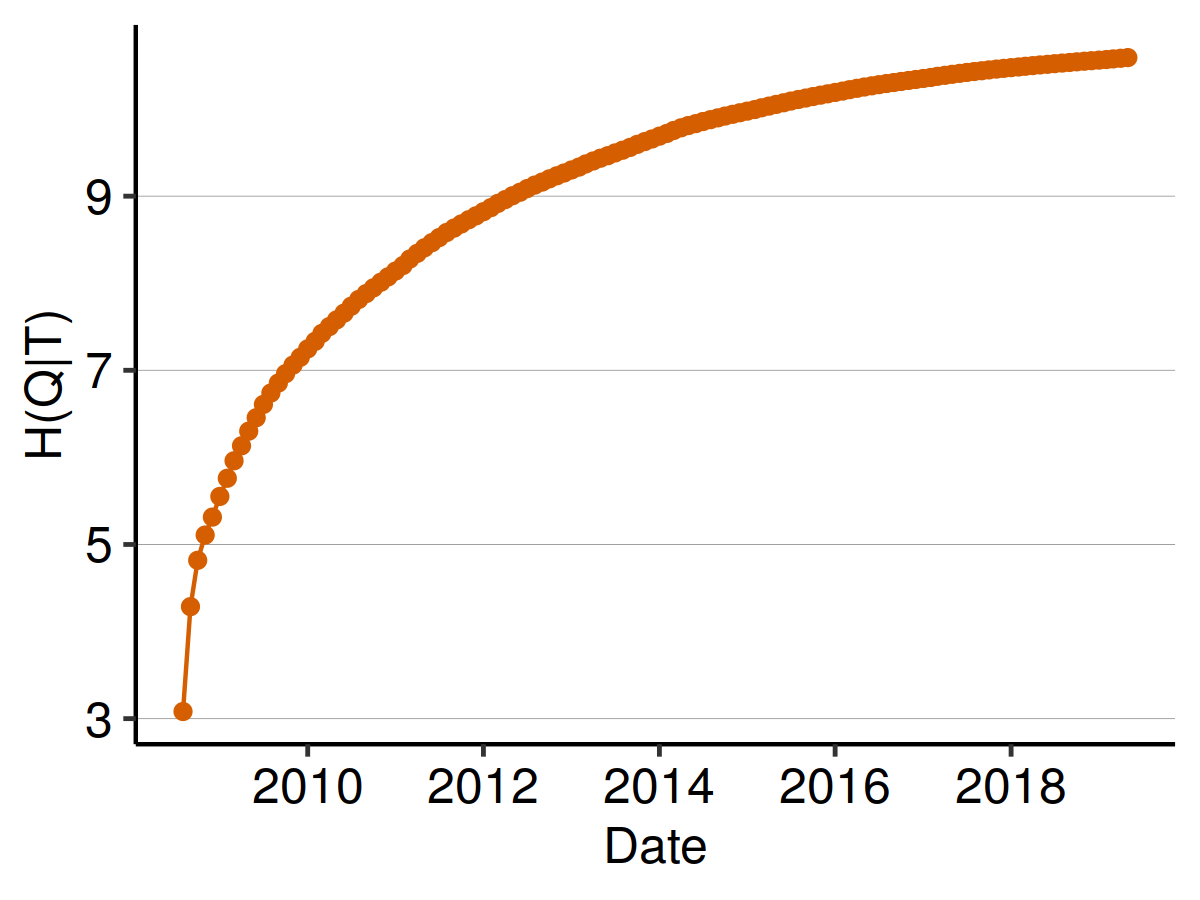}}
	\subfloat[Tag descriptiveness]{\includegraphics[width=0.24\textwidth]{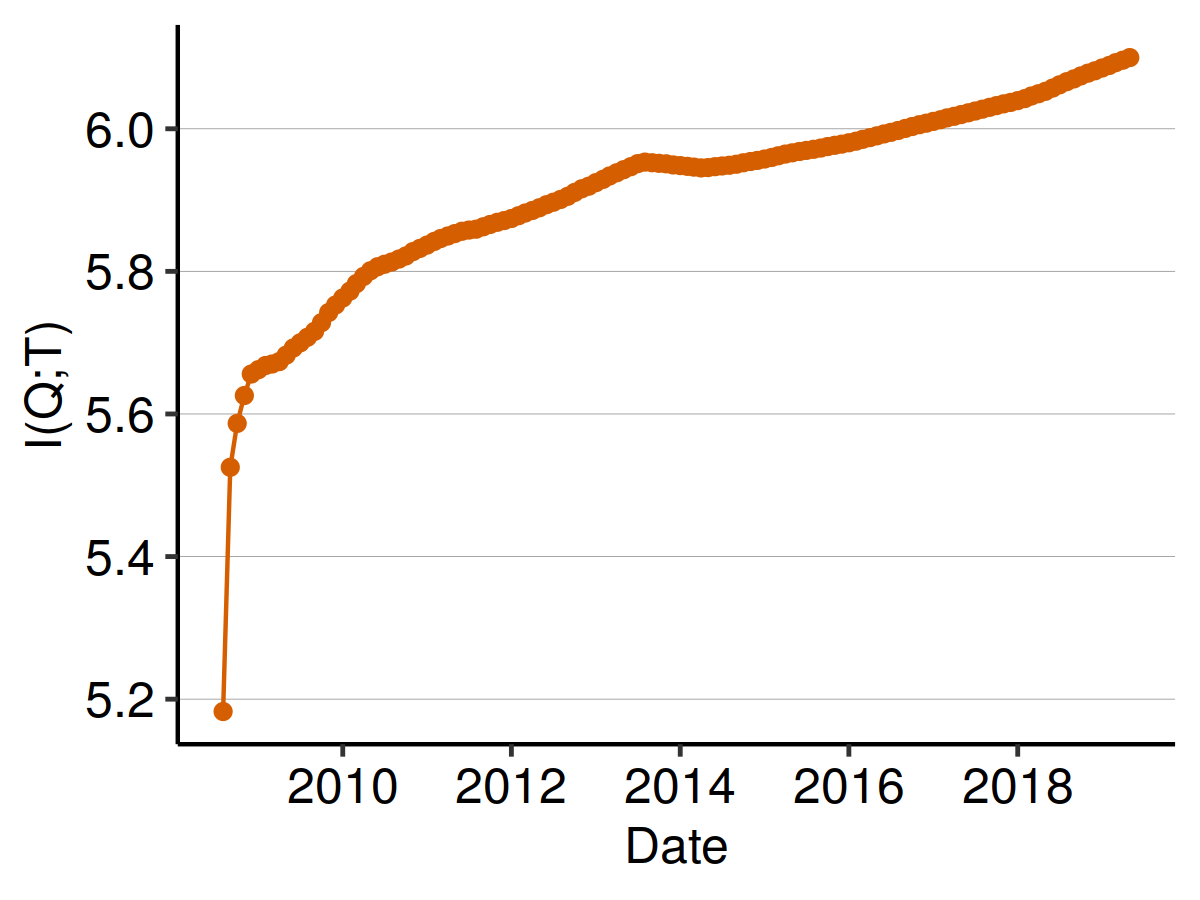}}
	\caption{
	\textit{Tag efficiency grows over time.}
		We compute the entropy of (a) questions and (b) tags as a measure of information content. Increasing entropy is better---the system has not saturated yet and every new question and assigned tags contribute to information content. 
		We measure retrieval difficulty as the conditional entropy of questions given tags. Lower entropy implies easier question retrieval via tags.
		We observe in (c) that the retrieval difficulty stabilizes after an initial phase of rapid growth. 
		The tag descriptiveness is captured with the mutual information between tags and questions. Higher values imply better descriptiveness.
		We observe in (d) that mutual information grows over time. All metrics indicate increasing tag efficiency on Stack Overflow, in contrast to del.icio.us~\cite{chi2008understanding}.
}
	\label{fig:retr_desc}
\end{figure*}

\section{Tag Efficiency on Stack Overflow}

\noindent \textbf{Information content.}
We observed in Figure~\ref{fig:data} that there is a pronounced growth in questions and tags over time, with a linearly increasing growth in recent years. 
The tag and question entropy gives insight in whether this steady growth of tags and questions also translates into growth of information content. 

We find that the entropy of questions and of tags increase throughout the observation period, as shown in  Figure~\ref{fig:retr_desc}a--b. 
This behavior is expected for questions as 
the entropy is $log|\mathcal{Q}|$. Growing $H(T)$ indicates that users contribute additional information when they assign tags to their questions, either by creating novel tags or by assigning more diverse and unpopular tags, which both lead to a less skewed tag distribution over time. Increasing question and tag entropy are desirable in a social tagging system, as growing information content is evidence of a healthy user community making significant contributions to the system.

\noindent \textbf{Information retrieval.}
In Figure~\ref{fig:retr_desc}c, we show the conditional entropy of questions given tags, Eq.~\ref{eq:condent}, as a measure of the difficulty to retrieve specific content using tags. The conditional entropy increases steadily, therefore content becomes increasingly difficult to retrieve. However, this trend subsides after 2016, signaling that 
the difficulty of retrieval begins to stabilize, despite increasing entropy. 
Notably, this result contrasts with measurements performed in previous work, which observed a continuous decrease in tag retrieval efficiency on a social bookmarking site called del.icio.us~\cite{chi2008understanding}. 

\noindent \textbf{Tag descriptiveness.}
Finally, we study the mutual information of questions and tags, Eq.~\ref{eq:mutual_info}, i.e., how well tags describe questions. 
In Figure~\ref{fig:retr_desc}d, we observe that $I(Q;T)$ steadily increases over time: thus tags tell us increasingly more about the corresponding questions.
Again, this observation stands in contrast to previous analysis of social tagging that reported declining tag descriptiveness when the number of tags is unconstrained~\cite{chi2008understanding}.

\begin{figure}[t]
	\centering
	\subfloat[New tag rate]{\includegraphics[width=0.24\textwidth]{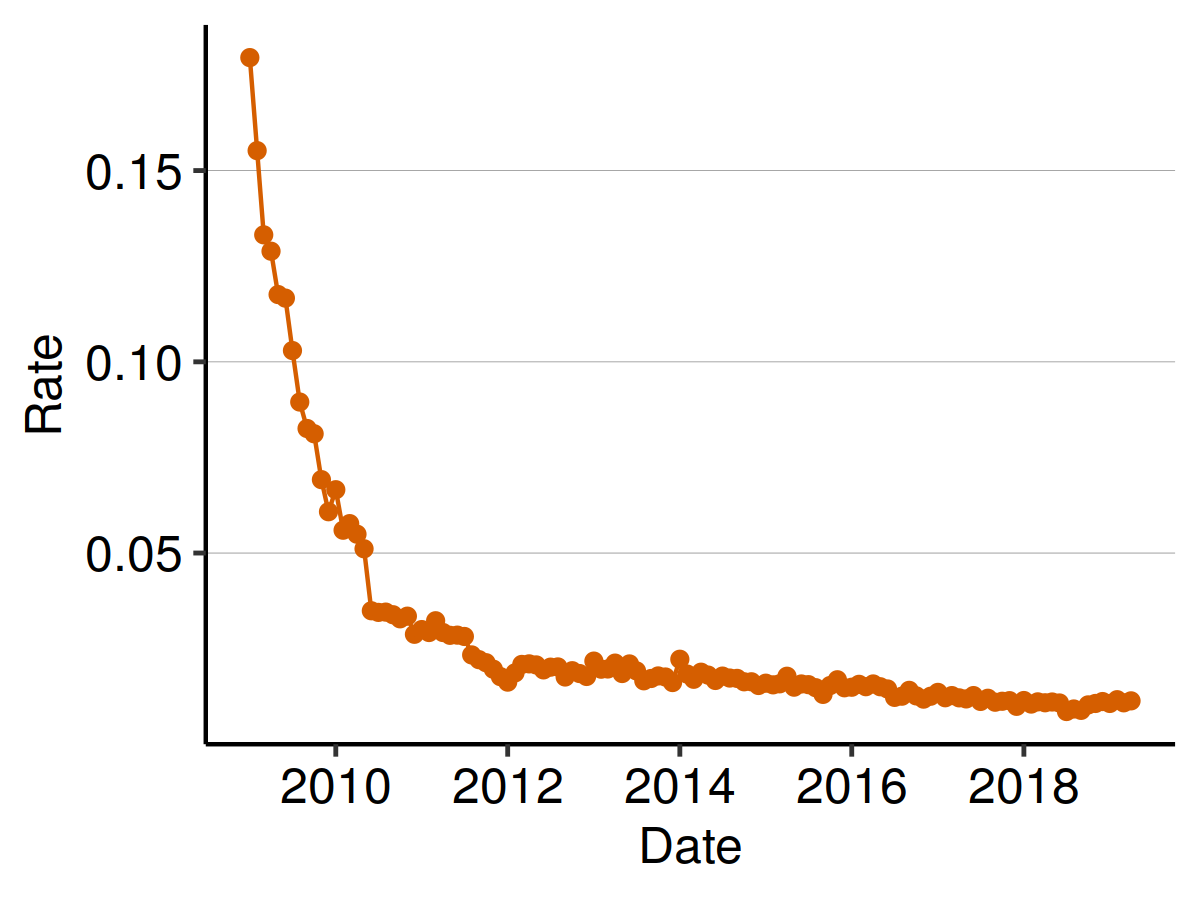}}
	\subfloat[Tag Gini coefficient]{\includegraphics[width=0.24\textwidth]{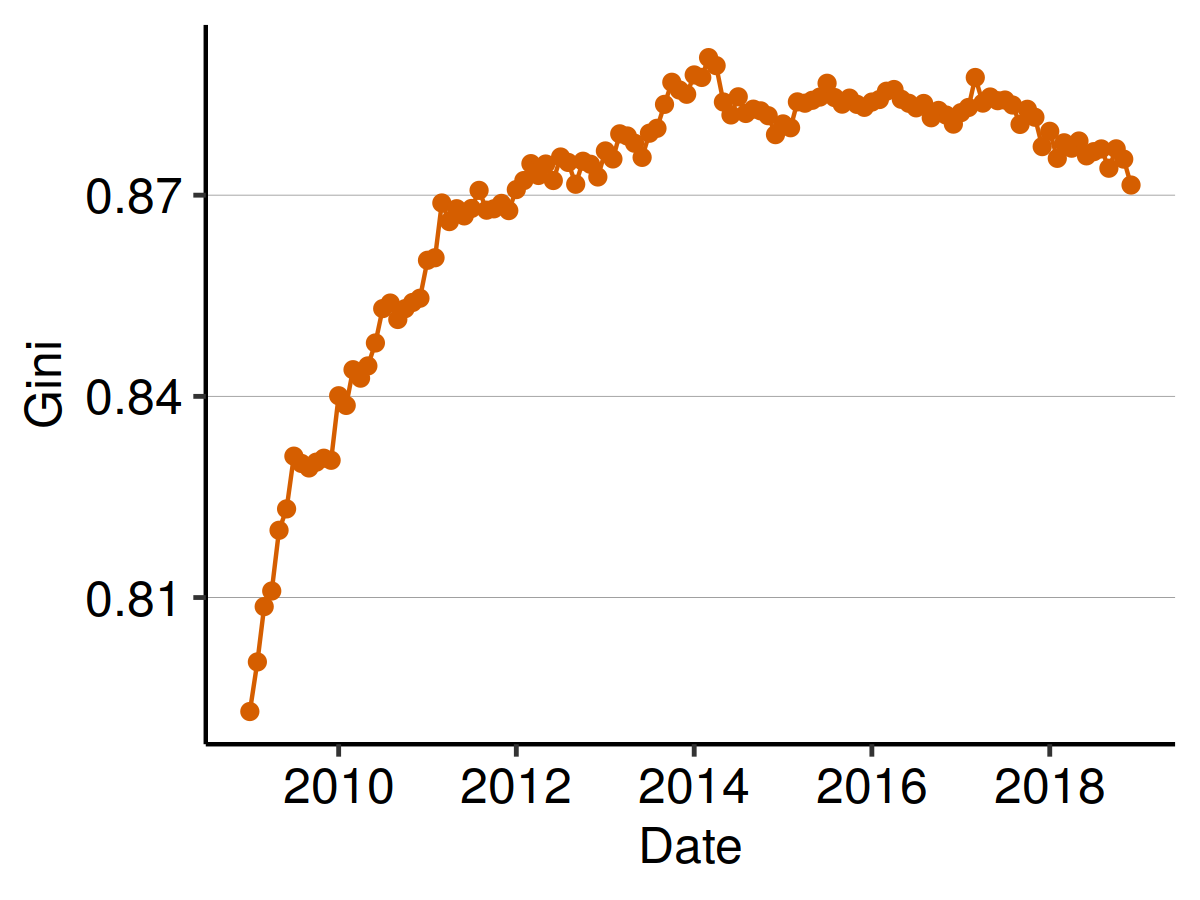}}
	\subfloat[Mean tag length]{\includegraphics[width=0.24\textwidth]{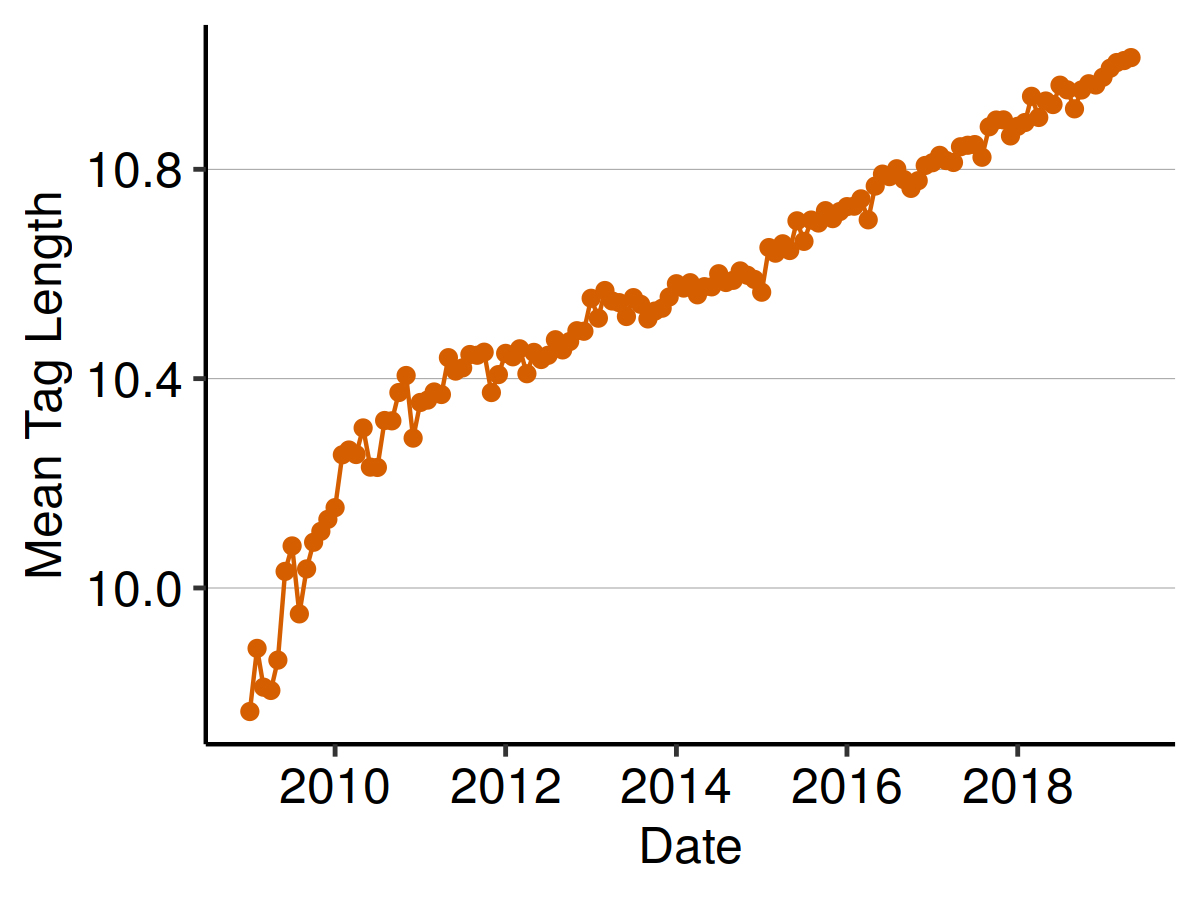}}
	\subfloat[Fraction of composite tags]{\includegraphics[width=0.24\textwidth]{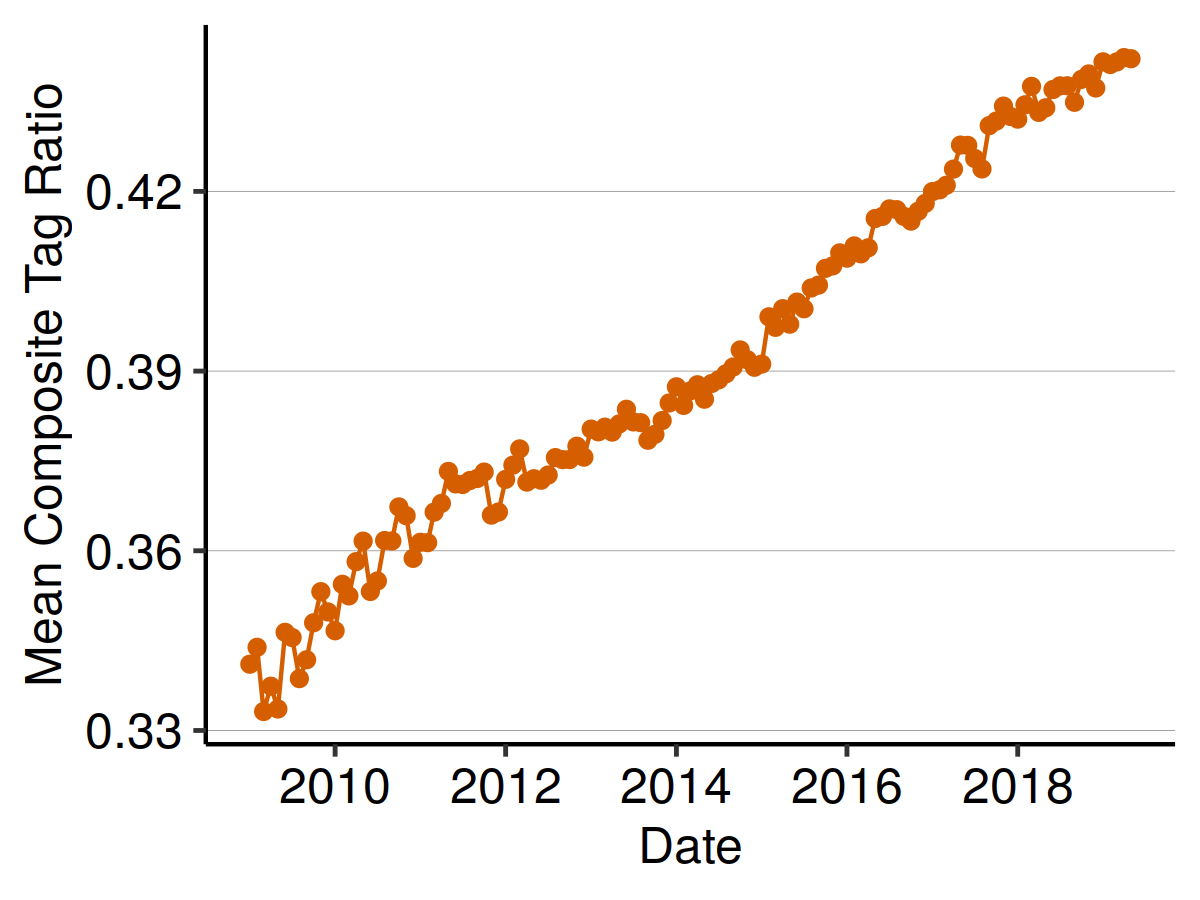}}
	\caption{\textit{Novelty, diversity, and specificity on Stack Overflow.} (a) The fraction of new tags per month decreases but stabilizes over time indicating steady arrival of novel tags. (b) Albeit the Gini coefficient of tags indicates popularity bias, tag inequality decays after 2014 signaling a steady increase in tag diversity. Assessing specificity as the tags' textual length and the fraction of composite tags (i.e., multiple words separated by a hyphen, such as ``google-cloud-firestore''), we observe in (c) and (d) that both measures grow over time. We hypothesize that increasing length of tags and growing adoption of composite tags may indicate increasingly specific tagging.}
	\label{fig:novelty_diversity}
\end{figure}

\noindent \textbf{Novelty and diversity lead to tag efficiency.}
To shed light on growing tag efficiency on Stack Overflow we analyze how our information-theoretic measures co-evolve in time. Question entropy $H(Q)$ on Stack Overflow is given by $log|\mathcal{Q}|$ due to the uniform question distribution. At the same time, tag descriptiveness given by $I(Q;T) = H(Q) - H(Q|T)$ grows steadily, meaning that $H(Q)$ has a higher growth rate than conditional entropy $H(Q|T)$. We can interpret this result as follows: although each new question and its tags add novel information to the system, it becomes easier to find the relevant information as users become more precise in describing questions with tags. This indicates an increasing tags specificity either through a more diverse usage of tags, creation of novel tags, or both. 
We corroborate this interpretation by looking at mutual information via tags: $I(Q;T) = H(T) - H(T|Q)$. The term $H(T|Q)$ gives the uncertainty of selecting a tag after selecting a question. 
As tags can not be repeated for individual questions on Stack Overflow this quantity is close to $log(3)$ (the average number of tags per question On Stack Overflow is around three, cf. Figure~\ref{fig:data}b). Hence, in $I(Q;T) = H(T) - H(T|Q)$, $H(T|Q)$ can be taken as constant and the growth of mutual information is reflected solely through the growth of $H(T)$, cf. the same shapes of $H(T)$ and $I(Q;T)$ in Figure~\ref{fig:retr_desc}b \&~\ref{fig:retr_desc}d again leading to the conclusion that the users on Stack Overflow are guided by \textit{novelty and diversity} in the tag usage.

We find empirical evidence for these theoretical considerations in data. Figures~\ref{fig:novelty_diversity}a--b show the monthly rate of new tags on Stack Overflow and the Gini coefficient as a measure of inequality of the tag distribution. We observe a decaying but a substantial tag novelty rate even eleven years after the system inception (Figure~\ref{fig:novelty_diversity}a), and an increasingly diverse usage of tags as we observe a downwards trend in inequality in recent years (Figure~\ref{fig:novelty_diversity}b). The starting value and growth of the Gini coefficient before 2014 indicates the presence of a strong tag popularity bias in the initial phase of Stack Overflow, which continuously subsides afterwards.

\noindent \textbf{Tags specificity and composite tags.}
Growing levels of tag descriptiveness in the long-term indicate that users become more specific and precise in their tag usage with time. Hence, we now investigate tag specificity as another mechanism that may play a key role in an increasing tag efficiency. Stack Overflow supports a kind of tag hierarchy, where a more general tag may form a superset for more specific tags.
For example, there is a tag termed ``google-chrome'' for general questions related to that Web browser. Further, there are also other, more specific tags on that topic, such as ``google-chrome-devtools'' or ``google-chrome-extension''.
We term such tags \textit{composite tags}, and we operationalize their usage as an indicator for the specificity of tags being applied to questions on Stack Overflow.
We measure the specificity of the tags (by computing the mean tag length in characters) and the usage of composite tags (by computing the mean monthly relative frequency of tags with at least one dash ``-''\footnote{Note that this is a noisy indicator for the usage of tag hierarchy, as the dash may also sometimes represent spaces in a tag's name, such as in the tag ``floating-point''. Manually annotating a set of $100$ randomly chosen composite tags, we find that $70$ leverage the dash to represent tag hierarchy rather than a space in the tag name. This suggests that a clear majority of tag names convey a tag hierarchy.}) over time in Figure~\ref{fig:novelty_diversity}c and Figure~\ref{fig:novelty_diversity}d, respectively. 

We find that tags become longer and use of composite tags increases, which may indicate growing tag specificity.
This observation may indicate stable tag retrieval efficiency in the long-term: the use of specific tags may help to uniquely identify questions especially given the tag limit, and thereby keep the uncertainty in question retrieval via tags stable---a result which was also found in previous work~\cite{Mamykina2011}. 

\section{Discussion}

\noindent \textbf{Single vs. multiple tags.}
Trying to replicate the finding that information content, tag retrieval efficiency, and descriptiveness decrease over time in tagging systems~\cite{chi2008understanding}, we instead find that on Stack Overflow, a system with comparable growth dynamics, information content and tag descriptiveness steadily grow, whereas retrieval efficiency stabilizes.
While the system analyzed in previous work~\cite{chi2008understanding}, a social bookmarking system called del.icio.us, remarkably differs in purpose and scope from the Q\&A system we study, we observe comparable activity dynamics but diverging developments in the organization of content.
The first substantial difference between the two systems is related to the multiple tagging of resources (Web pages on del.icio.us and questions on Stack Overflow). On del.icio.us multiple users can tag the same Web page, which potentially increases the skew in the distribution of resources (resulting in a less than maximal resources entropy). On contrary, on Stack Overflow every user creates a new question and assigns the tags at the time of creation. Hence, there is no reinforcement of questions as no other user can assign additional tags to an already existing question. This leads to a uniform question distribution and the maximal question entropy.

\noindent \textbf{Effects of tag limit.}
Apart from the dynamics of resource creation and tagging, another reason for the discrepancy between different social tagging systems may lie in the per-document \textit{tag limit}, which Stack Overflow imposes but del.icio.us does not. We believe that limiting the number of tags encourages users to create more specific tags (e.g.,  composite tags on Stack Overflow), which in turn weakens popularity bias as users assign more novel and diverse tags.

\begin{figure*}[t]
	\centering
	\subfloat[Tag limits vs. inf. retrieval]{\includegraphics[width=0.24\textwidth]{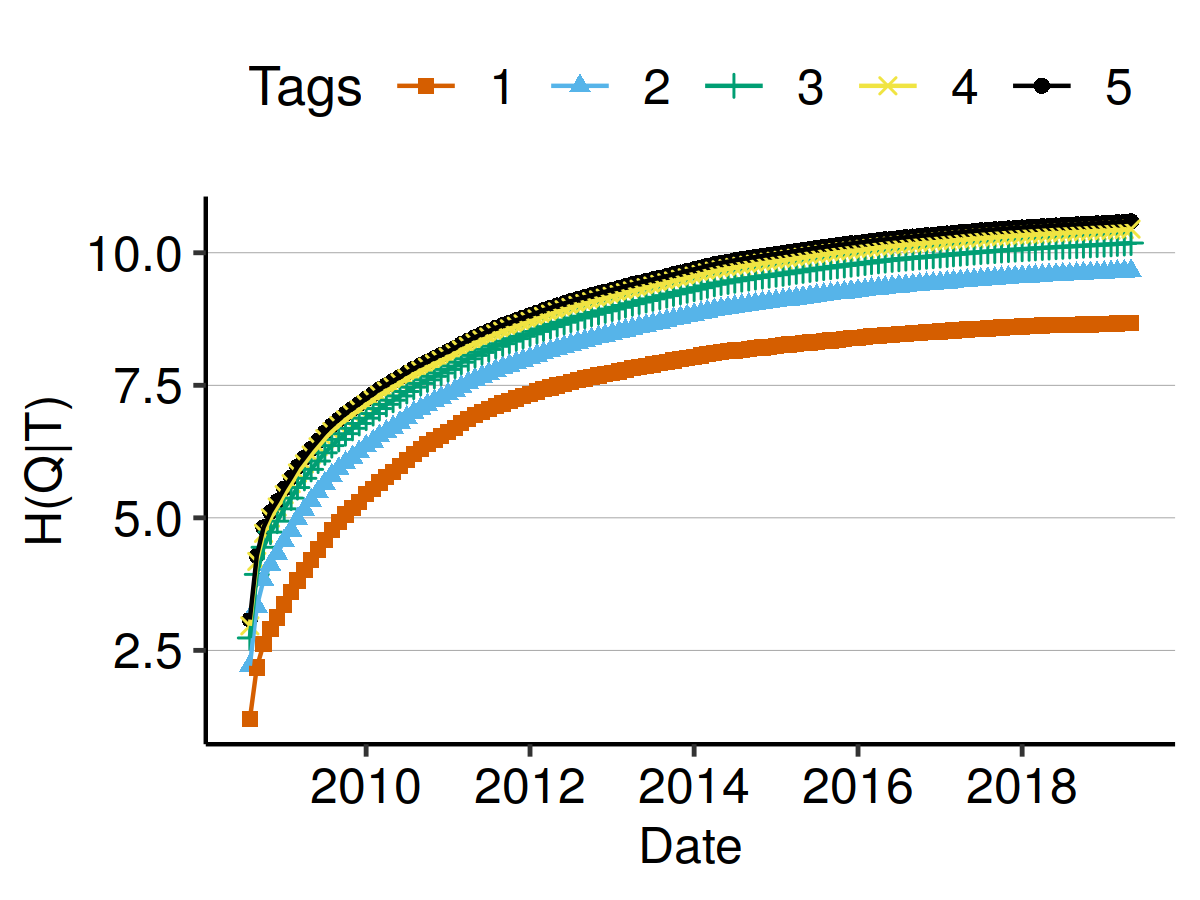}}
	\subfloat[Inf. content (tags)]{\includegraphics[width=0.24\textwidth]{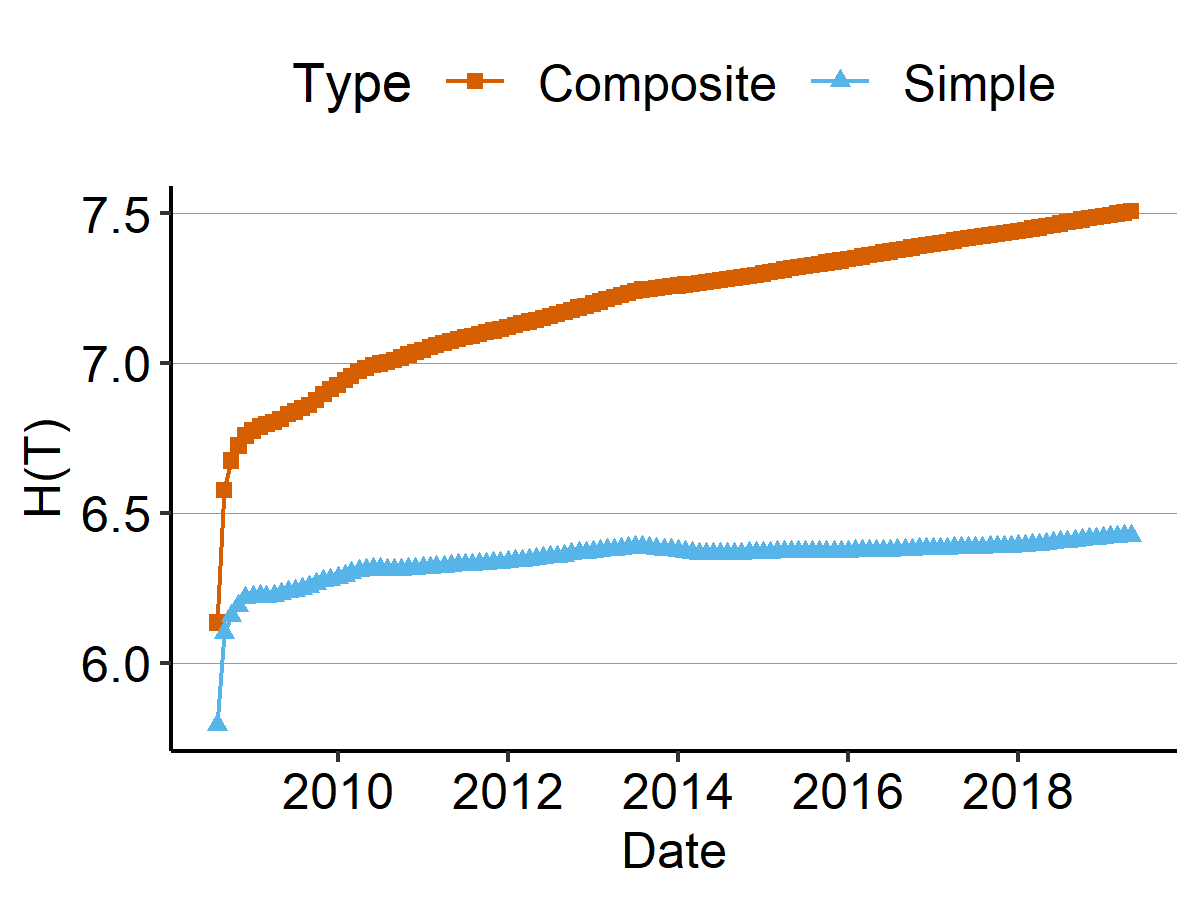}}
	\subfloat[Difficulty of inf. retrieval]{\includegraphics[width=0.24\textwidth]{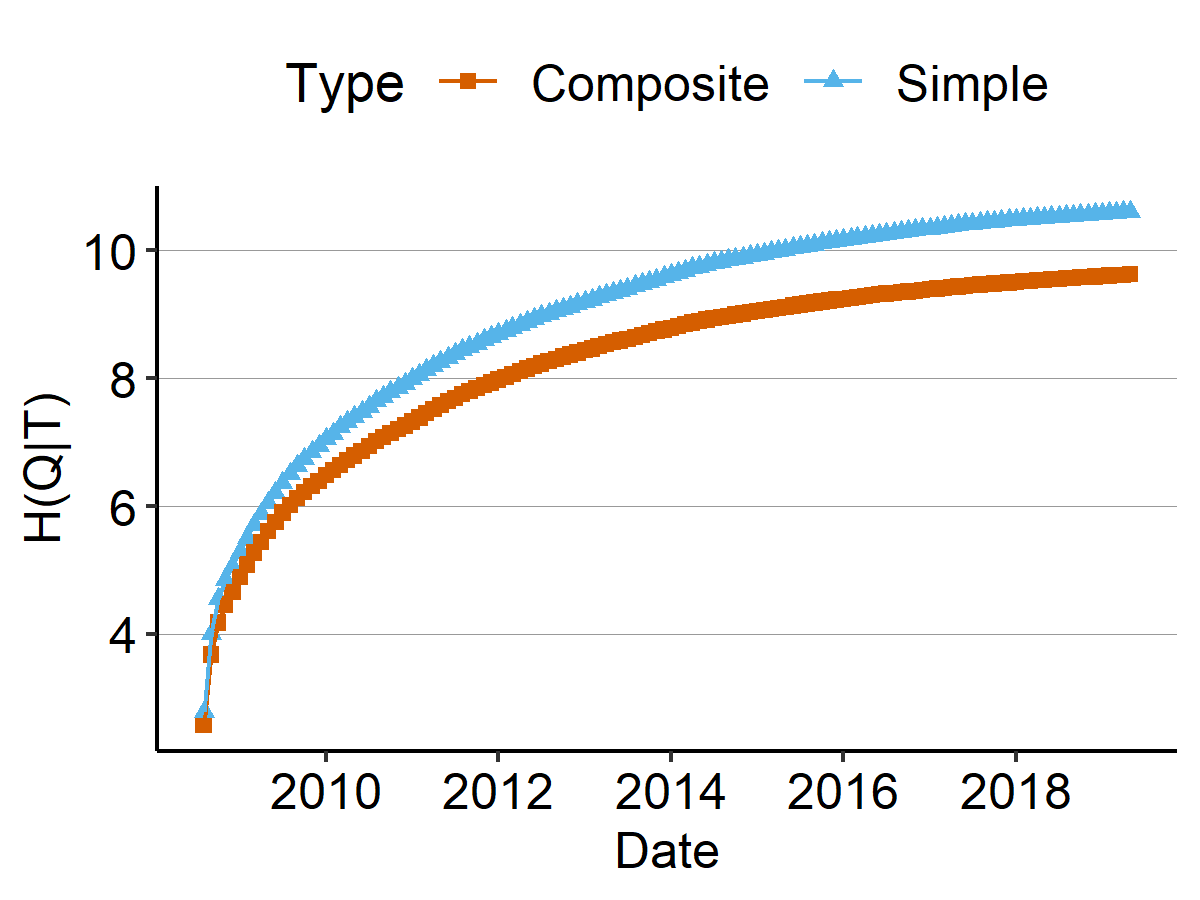}}
	\subfloat[Tag descriptiveness]{\includegraphics[width=0.24\textwidth]{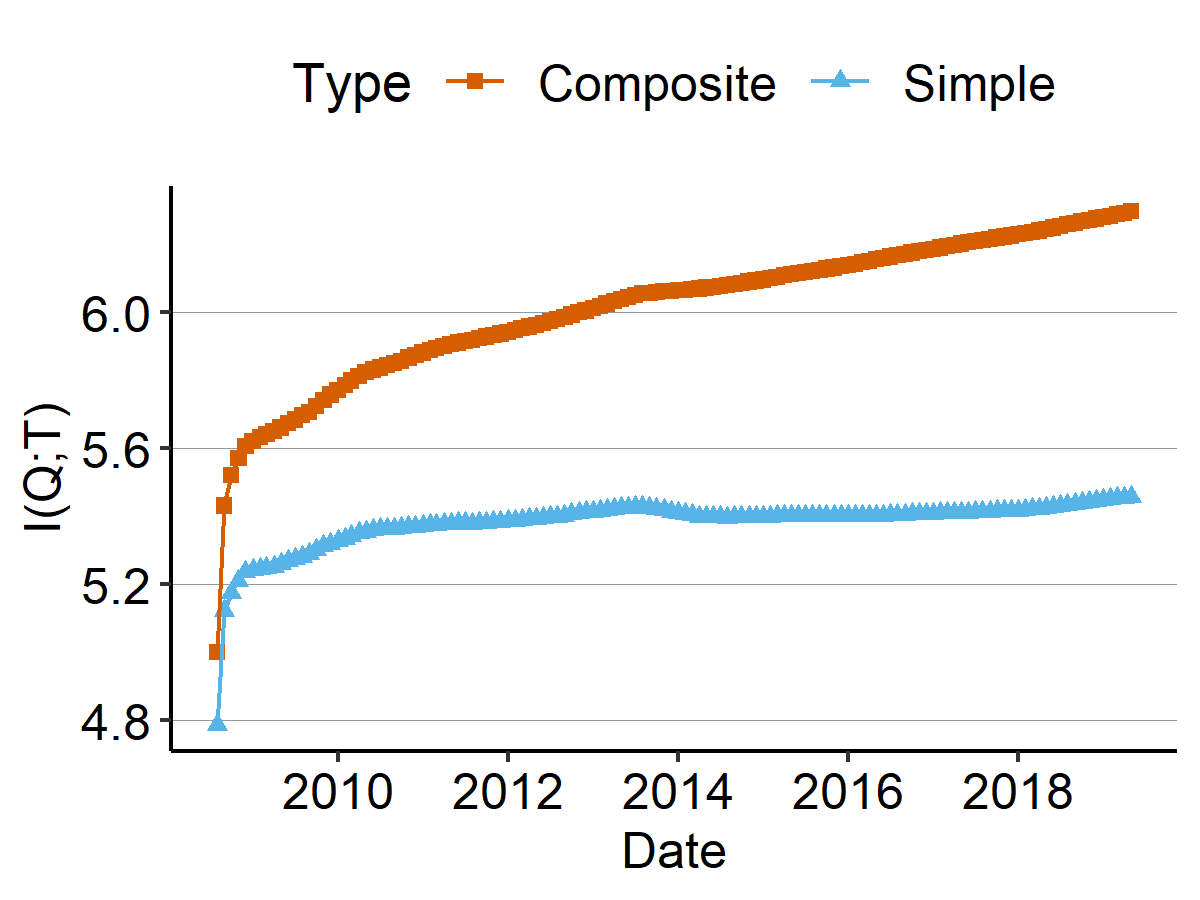}}
	\caption{\textit{Stricter tag limits and composite tags improve tag efficiency.} 
		We break down our measurement of the tag efficiency by questions with up to $x$ tags for $x=1,2,3,4,5$. Lower tag limits feature higher tag efficiency. Further, we observe in (a) a nonlinear, gradual decrease in tag efficiency as the tag limit increases.  
		To compare composite with simple tags we compute the tag entropy (b), conditional entropy of questions given tags (c), and mutual information (d) for questions with at least one composite tag vs. those without. 
		We observe that composite tags are more efficient than simple non-composite tags.}
	\label{fig:explanation}
\end{figure*}

To understand the impact of this design choice, we inspect the tag retrieval efficiency (cf. Eq.~\ref{eq:condent}) for different numbers of tags, as a proxy for the effects of imposing stricter limits on the number of tags per question. We plot the results of this exercise in Figure~\ref{fig:explanation}a.
We observe the lowest conditional entropy, i.e., highest tag efficiency for the strictest limit of one tag per question.
As the upper tag limit increases, the tag efficiency also gradually decreases to the level of questions with up to five tags. 
We observe that this gradual decrease is nonlinear: the difference in tag efficiency between the one and two tag limit is larger than between two and three tags, and this difference becomes successively smaller as the tag limit increases further.
These observations suggest setting a tag limit nudges users to employ specific, descriptive tags~\cite{koerner2010}, which in turn help organize and curate information.

\noindent \textbf{Effects of composite tags.}
In addition to the tag limit, simplicity of concatenating several tags with increasing specificity potentially accelerates the usage of composite tags (regardless of the tag limit) and contributes to tag efficiency on Stack Overflow. We analyze this effect by measuring, again via Eqs.~\ref{eq:ent_Q},~\ref{eq:condent}, and~\ref{eq:mutual_info}, the tag content, tag retrieval efficiency, and tag descriptiveness of questions with at least one composite tag vs. those without. We remark, with Figures~\ref{fig:explanation}b--d, that those with composite tags feature (a) higher tag content, also with a higher growth rate than the content of simple non-composite tags, (b) higher retrieval efficiency, and (c) higher tag descriptiveness again with a more pronounced growth as compared to simple tags.  
This observation validates the usage of composite tags as a solution which allows for specific tagging, and thus improved organization of information items~\cite{Mamykina2011}.
Nevertheless, on Stack Overflow the simple tags are also remarkably efficient as we observe increasing tag content, stabilizing retrieval efficiency and increasing tag descriptiveness of simple tags only (cf. blue lines in Figures~\ref{fig:explanation}b--d). This result is in stark contrast to del.icio.us~\cite{chi2008understanding} where all three metrics developed in the opposite directions. Hence, when controlling for the usage of composite tags, we still see a residual effect of growing tag efficiency on Stack Overflow, which we attribute to imposed tag limits and resulting increased levels of novelty and diversity even in the usage of simple tags.

A total disentanglement of the effects of the tag limit and other user interface choices is not possible: we can not decisively answer the question whether the tag limit or simplicity of creation leads to an increased use of composite tags and hence to a growing tag efficiency. We only observe that the combination of the tag limit and easy creation of composite tags results in novelty and diversity in tag usage and induces a strong positive effect on tag efficiency.

\noindent \textbf{Reducing popularity bias.}
Finally, we also reason about another feature of Stack Overflow's tagging mechanisms that may have positively impacted information organization: as users type the name of a tag for a question, an auto-complete system suggests pre-existing tags.
Crucially, though Stack Overflow currently sorts suggested tags by number of questions, this auto-complete system is not a tag recommendation system. As such, users are expected to provide descriptive tags on their own, and not rely on tag recommendations which may induce a stronger popularity bias~\cite{chi2008understanding}.

\subsection{Tag Growth Model}
To gain a further insight into our results, we present a model of social tagging that qualitatively reproduces the range of contrasting behaviors observed on Stack Overflow and del.icio.us~\cite{chi2008understanding}. 
Similarly to ´the models of Web growth~\cite{kleinberg1999} and the  evolution of social or technological systems~\cite{tria2013}, we base our model on an intuition that tagging decisions are correlated over time. In particular, our model is based on the mechanism of reinforcement (users copy decision of other users), novelty (users introduce new tags), and diversity (users diversify over popular as well as niche tags). Note that the goal of our model is to qualitatively study the dynamics of tagging behavior given a particular configuration of the model parameters and not to assess how user interface choices may translate to a specific set of model parameters.

Theoretically, our model closely resembles Polya's urn model. The model considers an urn containing some number of balls of different colors. At each step a ball is drawn randomly from the urn and then placed back together with a new ball of the same color, hence reinforcing the selected color. The reinforcement mechanism of the Polya's urn model results in a rich-get-richer dynamics and leads to power law distributions of the balls in the urn~\cite{yule1925}. Various variants of this model have been used extensively to model the power law phenomena in physics and computer science~\cite{tria2013,newman2005,SIMKIN20111}. 

\noindent \textbf{Reinforcement-novelty model.}
In the reinforcement-novelty model we start with two urns, one for the resources and one for the tags. At each discrete time step $i$ a new user tags a resource. First, the user decides with probability $p$ whether she wants to \textit{reinforce} a decision previously made by another user or whether she wants to introduce a \textit{novel} tag with probability $1-p$. In the case of reinforcement, the user selects a ball from the resource urn, then puts this ball back together with another ball of the same kind into the resource urn. This reinforcement mechanism correlates the decision of the current user to the past decisions of other users, thus introducing a \textit{popularity bias} into user decisions. On the other hand, in the case of novelty the user simply creates a new resource and adds it to the urn.
After the user selects the resource, she then decides how many tags to assign to that resource. We model this using random variable representing a limit on tags. We use a binomial random variable $Binomial(n_t, p_t)$ with $n_t$ corresponding to the tag limit and $p_t$ such that $n_tp_t$ reflects the empirical mean of tags per resource from a tagging system. For each individual tag, the user adopts the same mechanism as in the case of resources. Thus, she first decides with probability $q$ whether to copy an already existing tag or to assign a novel tag with probability $1-q$. In the case of reinforcement, the user first draws a tag from the tag urn, then puts this tag and one more copy of it back into the urn. In the case of a novel tag, the user creates a new tag and puts it into the tag urn.
After the user assigns the last tag, we repeat this process for a given number of users $u$. Finally, we inspect the information-theoretic measures to quantify tag efficiency.

\noindent \textbf{Reinforcement-novelty-diversity model.}
To account for \textit{diversity} in selection of tags during the reinforcement phase, we adapt the reinforcement-novelty model by changing the ball selection process from the tag urn. In particular, instead of using the classical Polya's urn mechanics, which selects tags proportional to their relative frequencies, we first smooth the frequencies. For smoothing we chose the softmax function:
\begin{equation}
    \sigma(\textbf{f})_i = \frac{e^{f_i/d}}{\sum_{i=1}^{|\mathcal{T}|}e^{f_i/d}},
\end{equation}
where $f_i$ is the relative frequency of tag $i$, $|\mathcal{T}|$ is the current number of tags, and $d$ is the \textit{diversity factor}, which controls the  strength of smoothing. 
When $d\rightarrow\infty$ (strong diversity), then $1/d\rightarrow 0$ leading to $\sigma(\textbf{f})_i\rightarrow 1/|\mathcal{T}|$, i.e., a high level of diversity results in a uniform tag distribution. On the other hand, when $d\rightarrow 0$ (weak diversity), then $1/d\rightarrow\infty$ resulting in the softmax function with probability $1$ for the maximal relative frequency with all other probabilities being $0$ (i.e, the extreme case of the popularity bias).
Thus, we can also interpret the quantity $1/d$ as the popularity factor. Note that the smoothing towards uniform distribution always happens when the differences between individual relative frequencies are smaller than the diversity factor $d$.
As the differences between relative frequencies are always from $[0, 1]$ setting the diversity to e.g., $1$ achieves smoothing, whereas setting diversity to a value close to $0$ induces a strong popularity bias.

\begin{figure*}[t]
	\centering
	\subfloat[Reinforcement]{\includegraphics[width=0.25\textwidth]{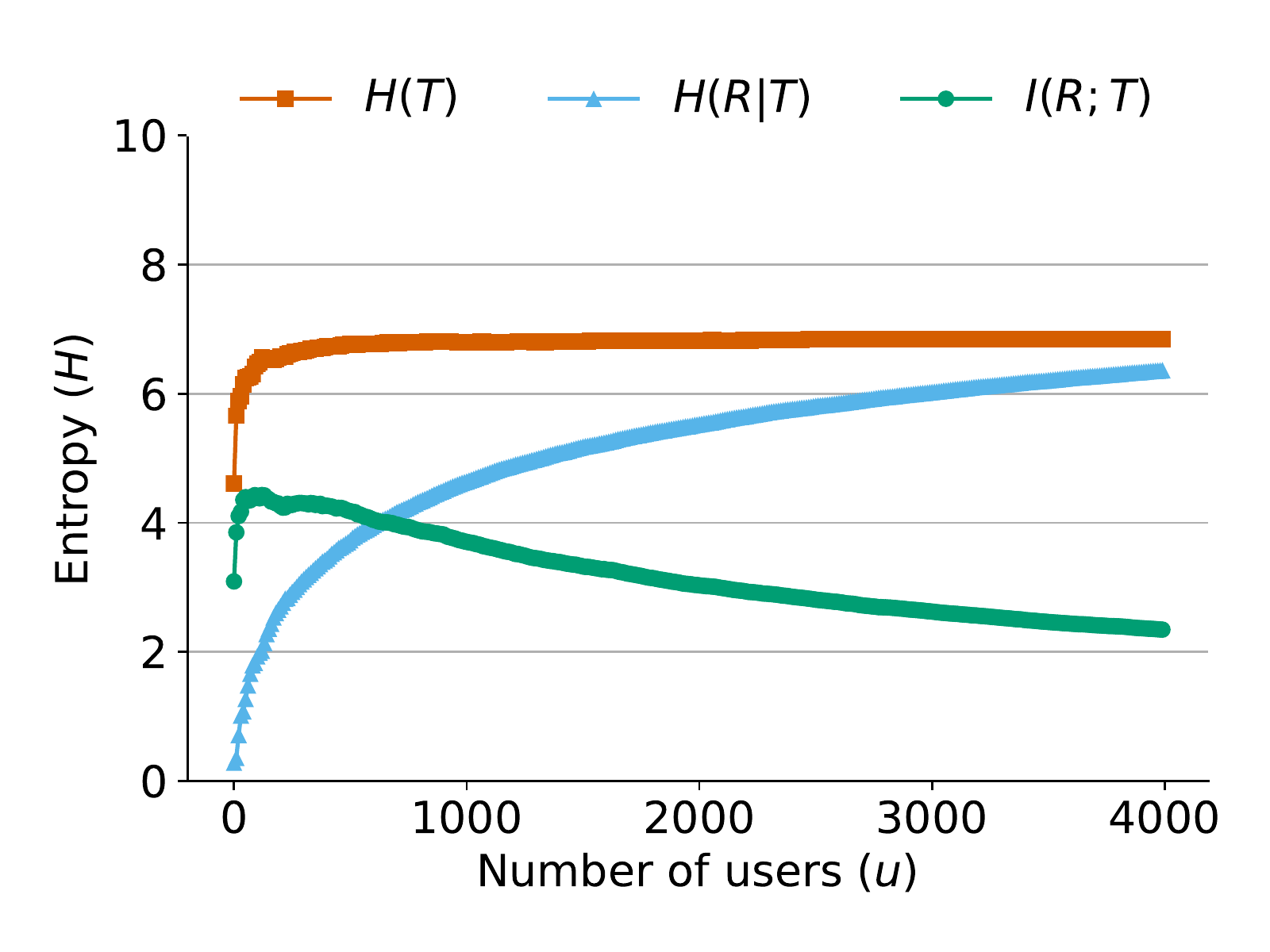}}
	\subfloat[Reinforcement-diversity]{\includegraphics[width=0.25\textwidth]{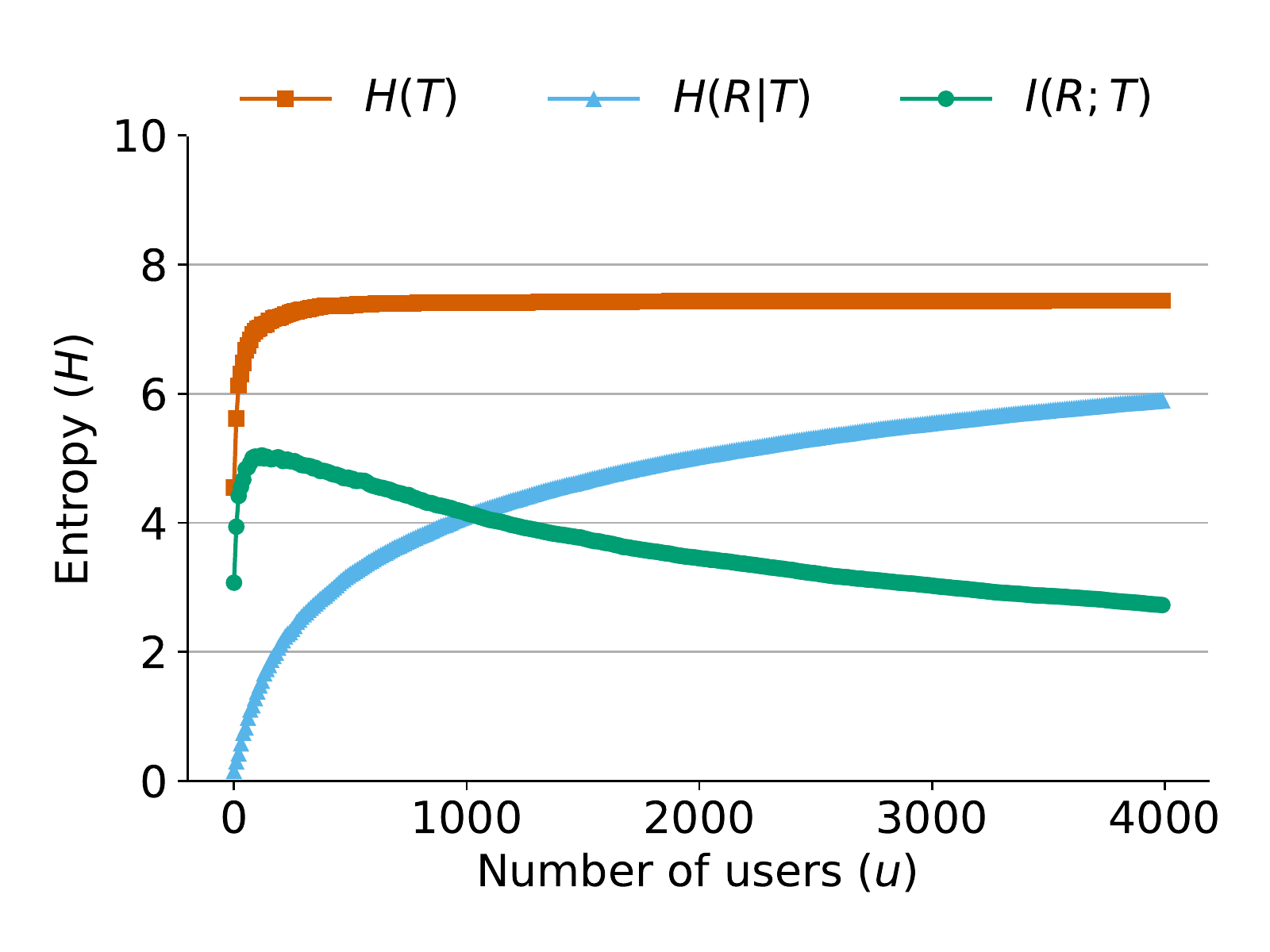}}
	\subfloat[Reinforcement-novelty]{\includegraphics[width=0.25\textwidth]{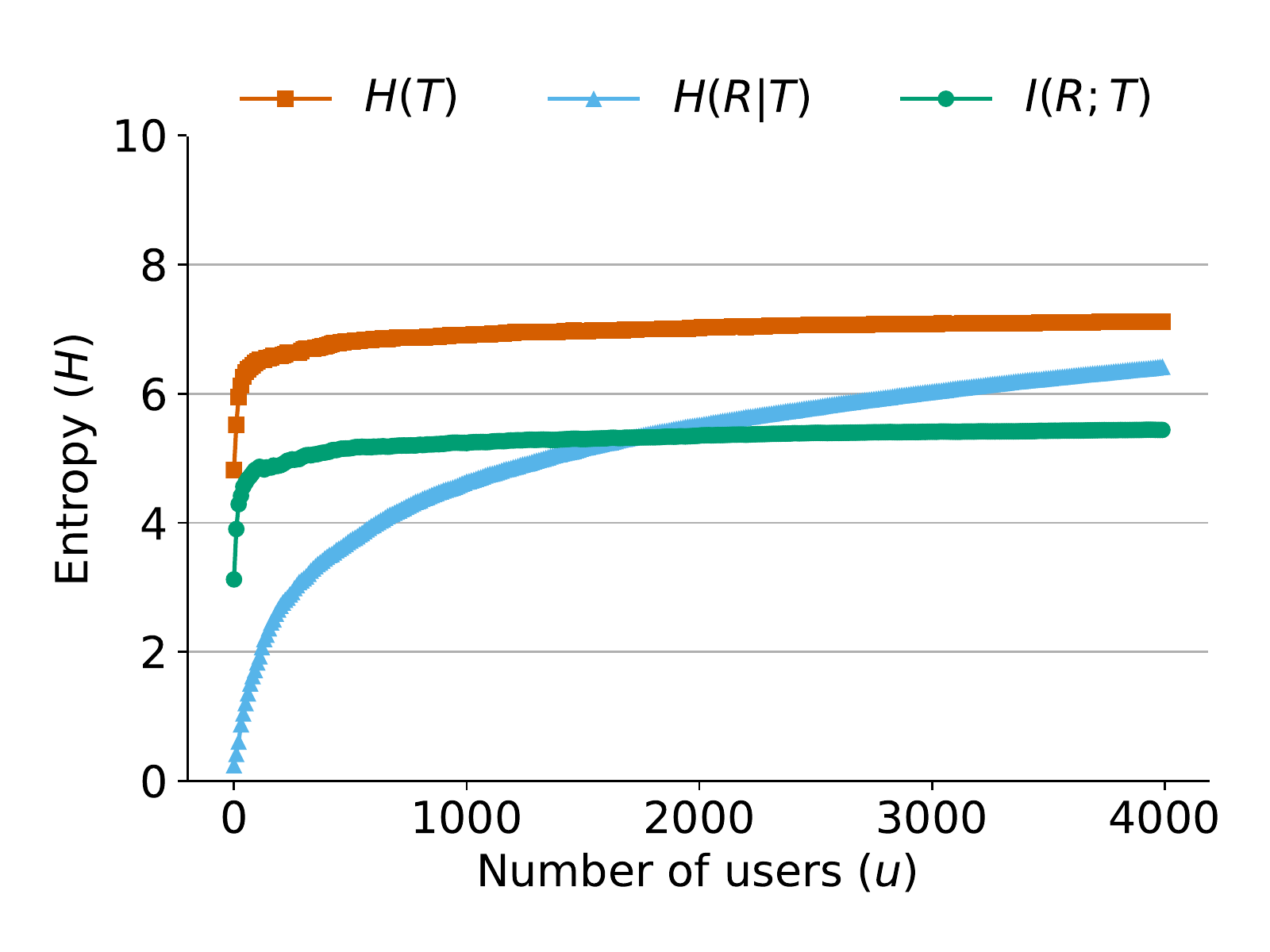}}
	\subfloat[Reinforcement-novelty-diversity]{\includegraphics[width=0.25\textwidth]{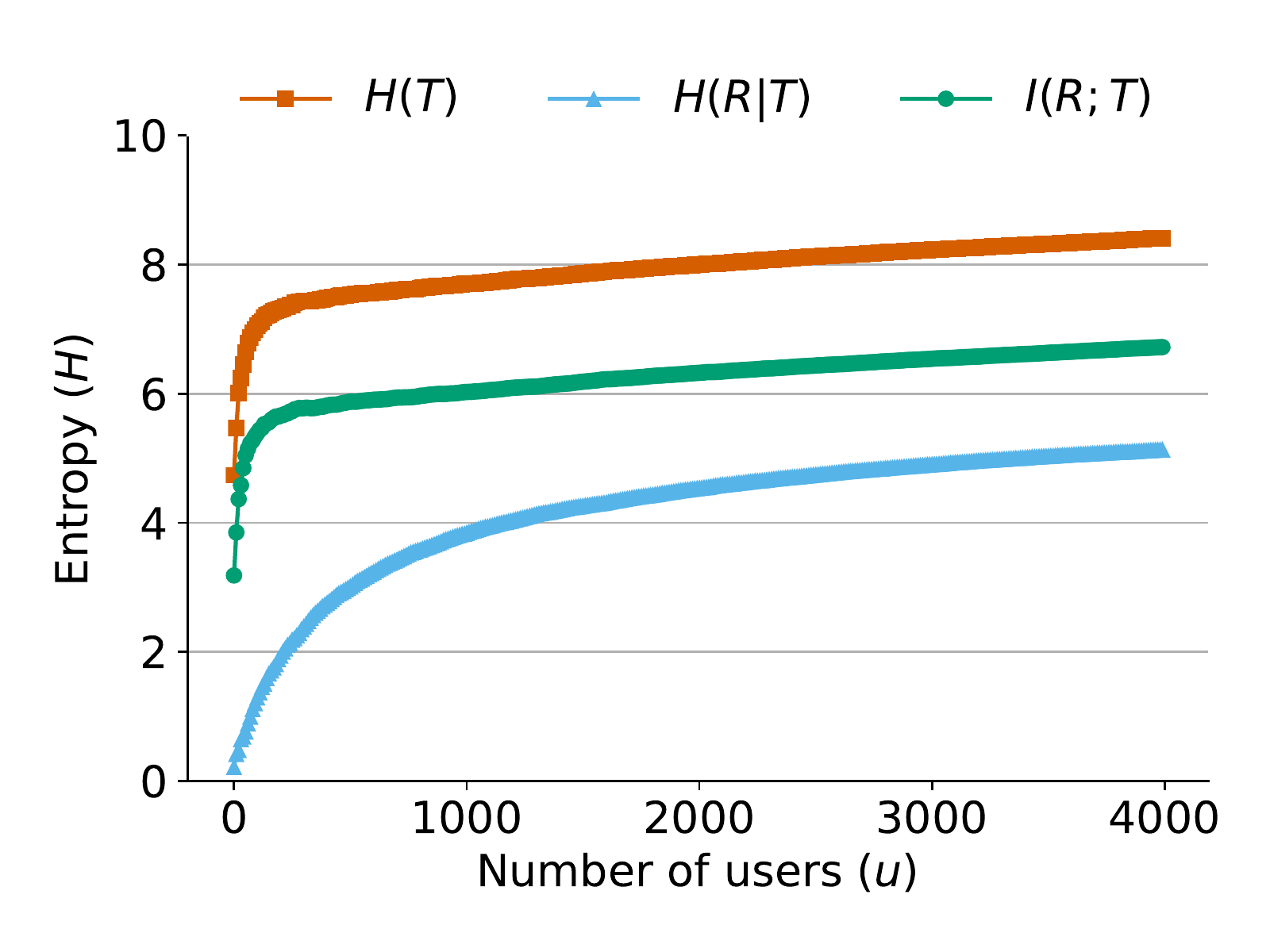}}
	\caption{\textit{A modeling of social tagging reproduces a range of behaviors observed in real systems.} (a) In a pure reinforcement system (i.e., $p=1$ and $q=1$) with a strong popularity bias, we observe a deteriorating tag performance across all measures. (b) In a system with diverse selection of tags, tag efficiency improves but still decays over time.
	(c) In a reinforcement-novelty system with $p=0$ (each user creates a new question) and $q=0.98$ (most recent observed rate of tag reinforcement in Stack Overflow) but without diversity, tag entropy and mutual information develop more positively, but the tag retrieval efficiency decays quickly. (d) Finally, in a system with reinforcement, novelty, and diversity (e.g. Stack Overflow), we observe increasing tag efficiency with a stabilizing conditional entropy.}
	\label{fig:model}
\end{figure*}

\noindent \textbf{Comparison with existing models.}
We briefly compare our model with two most prominent models of the growth of information systems:
\begin{inparaenum}[(a)]
\item the model of correlated novelties by Tria et al.~\cite{tria2013}, and
\item the copying model by Kleinberg~\cite{kleinberg1999}.
\end{inparaenum}
The first difference is that in our model we have two sequences, one for the resources and one for the tags. Two other models have only one sequence each, e.g., the Web pages in Klenberg's model, or Wikipedia pages in Tria's model. Another important difference between our reinforcement-novelty-diversity variant and the other two models is the smoothing mechanism, which we use to reduce the popularity bias. 
Further, in contrast to the model of correlated novelties, there are no explicit correlations in our model, neither within the urns nor between the urns,
although in actuality there will be  correlations when selecting tags for a given resource, e.g. due to  semantics. However, there are implicit correlations between popular resources and tags that are frequently selected together due to the reinforcement.
Finally, in the  correlated novelties model the reinforcement and  novelty rates are dynamically changing with time. The model follows Heap's law~\cite{heaps1978}, which implies that the novelty rate decreases with time as $t^{\beta-1}$, where $\beta$ is typically estimated from the data (e.g., $\beta=0.78$ for del.icio.us)~\cite{tria2013}. We estimate the Heap's law exponent for Stack Overflow as $\beta=0.57$ (distribution head fit) and $\beta=0.315$ (full data fit). Slightly lower values of the Heap's law exponent may be potentially caused by the limit on the number of tags per question and proposal of new tags on Stack Overflow.
We leave the extension of the model to dynamic novelty and diversity rates  for future work.

\noindent \textbf{Fitting model parameters.} We sketch here shortly how model parameters may be estimated from the data. Reasonable estimates for novelty $1-p$ and $1-q$ are fractions of newly introduced resources respectively tags as this is maximum likelihood estimator for the Bernoulli random variable. For diversity $d$, we could adopt a numerical optimization for computing maximum likelihood estimator for the product of the softmax functions.

\noindent \textbf{Simulations.}
We simulate the growth of the tagging system for $u=4000$ users (larger numbers create analogous effects). We draw the number of tag assignments per user from a $Binomial(5, 0.6)$ distribution with mean of $3$, which corresponds to the empirical mean of the number of tags per question on Stack Overflow (Fig.~\ref{fig:data}b). We then perform a series of experiments with various variants of the model including reinforcement-novelty, reinforcement-diversity, and reinforcement-novelty-diversity with different values for reinforcement probabilities $p$ and $q$, and diversity factor $d$. 

Figure~\ref{fig:model} shows simulation results. The model qualitatively reproduces the behavior of social tagging systems. In a reinforcement-only system (e.g., del.icio.us~\cite{chi2008understanding}) we observe that initial phase of increasing tag efficiency is followed by a sharp drop and a steady decay in all of three information-theoretic measures (Fig.~\ref{fig:model}a). Extending the model to include diversity
($d=1$), improves decay rate slightly but still decreases performance in the long-term (Fig.~\ref{fig:model}b). In the next model variant, we set $p=0$ to model question creation in Stack Overflow, where each user creates a new resource (question). Further, we set $q=0.98$ modelling weak novelty in tag creation but without diversity. The system performance improves as we observe the growth of tag entropy and a weak but steady increase in mutual information. Still, the conditional entropy
increases with the number of users, signaling a drop in the retrieval efficiency (Fig.~\ref{fig:model}c). Finally, we model a system with novelty ($p=0$, $q=0.98$) and diversity ($d=1$) and are able to recover the empirical results from Stack Overflow. There is a constant increase in tag entropy and tag descriptiveness and a weak rate of increase in the conditional entropy of questions given tags indicating a stabilizing tag retrieval performance ( Fig.~\ref{fig:model}d).

\begin{figure*}[t]
	\centering
	\subfloat[Reinforcement-novelty]{\includegraphics[width=0.45\textwidth]{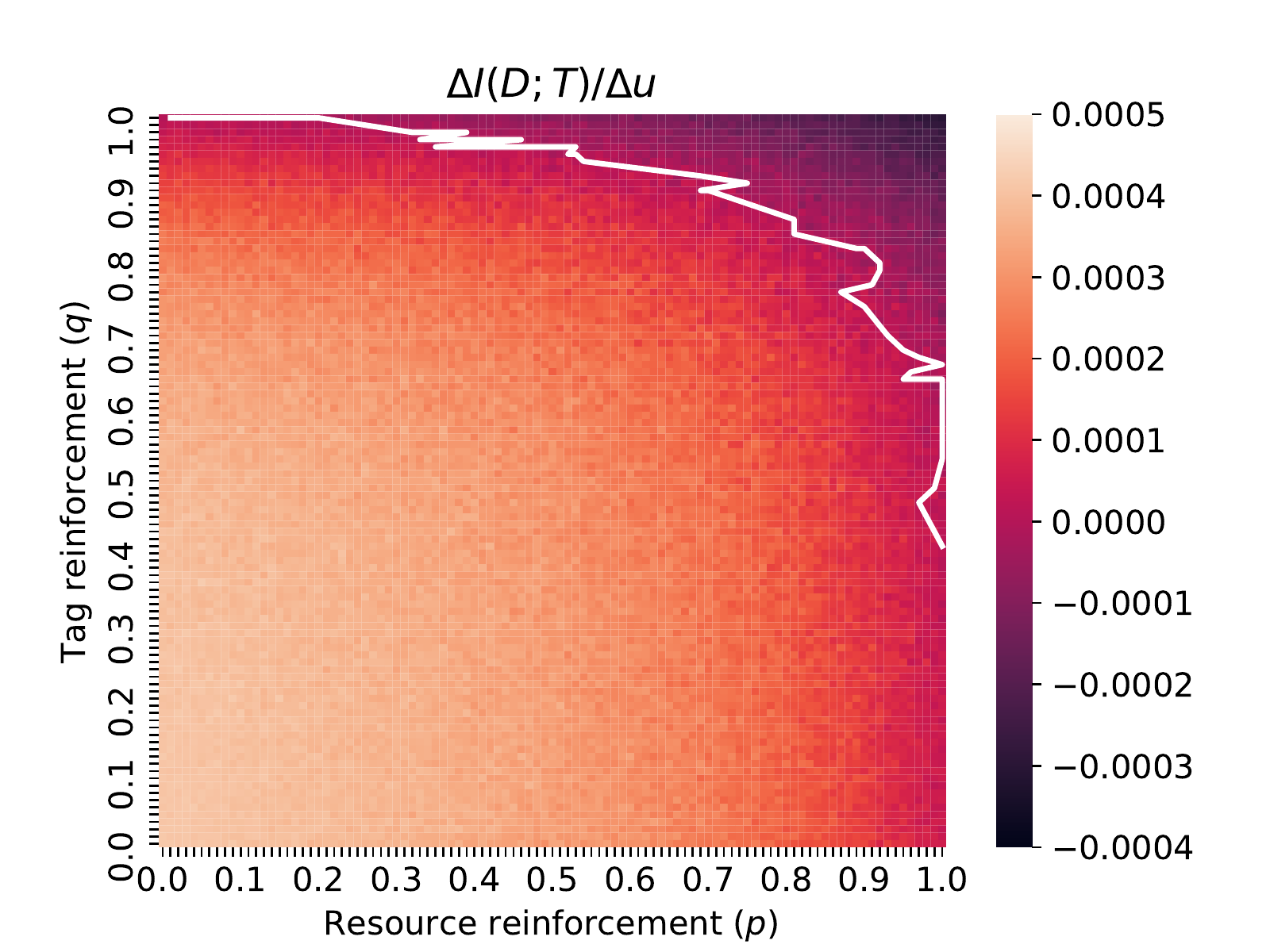}}
	\subfloat[Reinforcement-novelty-diversity]{\includegraphics[width=0.45\textwidth]{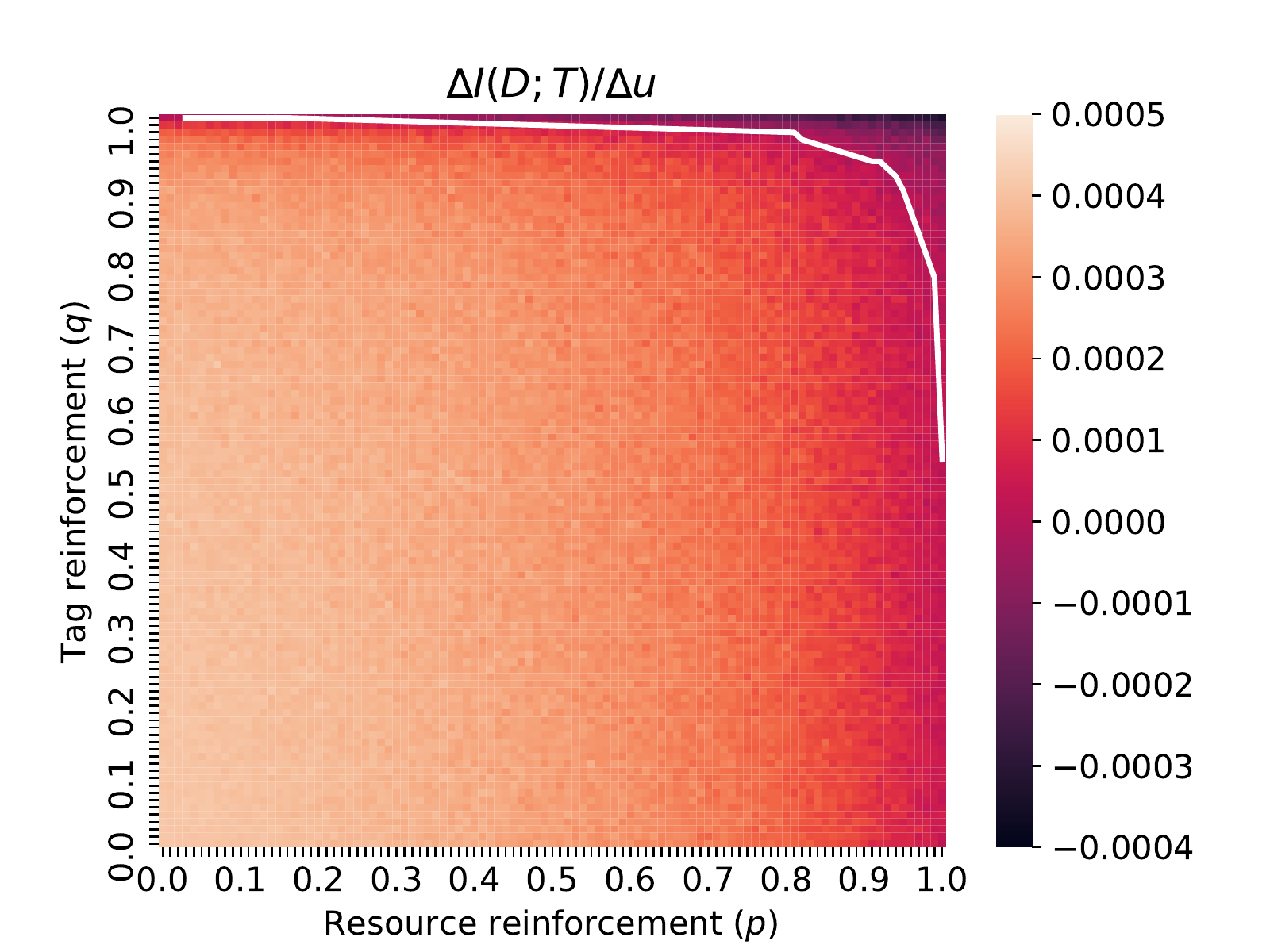}}
	\caption{\textit{Novelty and diversity lead to tag efficiency.} The heatmaps show the average rate of change of tag descriptiveness in the long-term. Below the white lines the rate of change is positive (similarly to Stack Overflow), above the lines negative (similarly to del.icio.us). The high levels of novelty in tag and resources usage lead to an efficient system, whereas stronger reinforcement of both tags and resources causes a steep drop in tag efficiency. Without diverse selection of tags (a), the performance drop is more pronounced, while diversity in the tag usage (b) increases robustness of the system with respect to tag efficiency.}
	\label{fig:rates}
\end{figure*}

In our next experiment, we iterate over the entire range $[0, 1]$ for both $p$ and $q$, keeping other parameters fixed. 
We measure the average rate of change of all three information-theoretic measures for the last $1000$ users, after the systems enter a stable phase. We repeat this experiment two times, the first time without diversity and the second time with diversity ($d=1$). Figure~\ref{fig:rates} shows the heatmaps of the rate of change of mutual information between resources and tags for those two experiments. In both experiments, we observe a radial decrease in the rate of change starting from the origin, which corresponds to a pure novelty system ($p=0$, $q=0$).
The rate of change is $0$ along the white lines, and is negative above the lines corresponding to a system of decaying tag performance, such as del.icio.us. At the point of pure reinforcement ($p=1$, $q=1$) the system performance is the worst and mutual information rate of change is most negative. We also observe a substantial difference between the model with and without diversity.
In the reinforcement-novelty-diversity model, the drop in the rate of change is much slower, indicating a robust tag performance as a consequence of a more diverse tag selection. To save space, we do not show the heatmaps for tag entropy and conditional entropy: both of which show similar deteriorating tag performance with increasing reinforcement probabilities. 

%
%--------------------------------------------------
%
\subsection{Limitations}
\label{sec:lim}
This work analyzes and helps explain the long-term organization of information in Stack Overflow. 
While we believe that the imposed tag limit relates to the development in tag efficiency, we do not claim a causal relationship between the limit in tags and the tag efficiency. 
Platform-level differences between the dataset studied in previous work~\cite{chi2008understanding} and Stack Overflow may confound the attribution of better tag efficiency in Stack Overflow to the tag limit ad composite tags. 
For example, on Stack Overflow only experienced users are allowed to create new tags---experienced users have a better understanding of the domain and a better overview of the tags that are still missing in the system and can, hence, introduce more specific and precise tags, which are beneficial for tag efficiency. Previous work has already suggested the importance of expert moderation for community sense-making via tags~\cite{Mamykina2011}. Another confounding factor may be the limited scope of Stack Overflow, which focuses on computer science and programming---users of Stack Overflow have typically computer science background and this supports a more efficient knowledge organization as compared to a more general topic.
Explicitly controlling for specific community characteristics may allow for a fairer comparison, and applying our analyses to other Stack Exchange topics and online communities would generalize our study.
Alternatively, deriving some kind of counter-factual world from the observational Stack Overflow dataset might help to understand the impact of the tag limit. In a counter-factual world, there would not be limits to tags and we would observe how tag efficiency evolves over time. As this is not possible (even for Stack Overflow managers), one approximation to such a setting may involve extracting and studying descriptive terms from the body of a question, which consists of free-form text and thus does not suffer from the same restriction as tags do.
Overall, we believe that there is an opportunity for future work to explore causal disentanglement of factors which may confound the impact of tag limits on tag efficiency.

Finally, our tag growth model can inform system operators qualitatively about the system evolution but can not produce exact quantitative predictions. By adopting a parameter fitting procedure as we sketched above, short-term predictions are possible---we leave the performance evaluation of such a predictive model for future work. Nevertheless, we see the utility of our model in possibility to explore other similar platforms.

\section{Conclusion}

In this work, we studied information growth and organization on the Q\&A website Stack Overflow. 
In particular, we inspected the effect of tag limits on the tag efficiency. 
To that end, we employed a set of information-theoretic metrics, which captured information growth and information organization. 
We found that, while information is growing in Stack Overflow, the decrease in tag efficiency stabilizes while tag descriptiveness increases, in contrast to previous results which indicated steadily declining tag efficiency~\cite{chi2008understanding}. 
Exploring alternative explanations for Stack Overflow's long-term stability in tag efficiency, our analyses uncover the importance of the tag limit and composite tags in supporting annotation of information while maintaining long-term tag efficiency.
Thus, this work indicates that system managers may trade-off between annotation capacity and tag efficiency with the choice of threshold for the tag limit per resource.

In the future, we will compare and extend our study to other Stack Exchange topics and Q\&A systems such as Yahoo! Answers, or even generalize it to systems with alternative information organization approaches, such as Wikipedia.
A study of these different approaches for information organization may be formulated within our information-theoretic framework. As such, extending our operationalization of information theory to learn more about information growth and organization poses an exciting prospect for future work. Furthermore, it will be critical to use a controlled experiment or causal analysis to infer the causal effect of tag limits and other designs to improve information organization.

\bibliographystyle{ACM-Reference-Format}
\bibliography{bib}


\begin{thebibliography}{47}

%%% the Web address.

\ifx \showCODEN    \undefined \def \showCODEN     #1{\unskip}     \fi
\ifx \showDOI      \undefined \def \showDOI       #1{#1}\fi
\ifx \showISBNx    \undefined \def \showISBNx     #1{\unskip}     \fi
\ifx \showISBNxiii \undefined \def \showISBNxiii  #1{\unskip}     \fi
\ifx \showISSN     \undefined \def \showISSN      #1{\unskip}     \fi
\ifx \showLCCN     \undefined \def \showLCCN      #1{\unskip}     \fi
\ifx \shownote     \undefined \def \shownote      #1{#1}          \fi
\ifx \showarticletitle \undefined \def \showarticletitle #1{#1}   \fi
\ifx \showURL      \undefined \def \showURL       {\relax}        \fi
% invisible to TeX
\providecommand\bibfield[2]{#2}
\providecommand\bibinfo[2]{#2}
\providecommand\natexlab[1]{#1}
\providecommand\showeprint[2][]{arXiv:#2}

\bibitem[\protect\citeauthoryear{Ahasanuzzaman, Asaduzzaman, Roy, and
  Schneider}{Ahasanuzzaman et~al\mbox{.}}{2016}]%
        {ahasanuzzaman2016mining}
\bibfield{author}{\bibinfo{person}{M. Ahasanuzzaman}, \bibinfo{person}{M.
  Asaduzzaman}, \bibinfo{person}{C.K. Roy}, {and} \bibinfo{person}{K.A.
  Schneider}.} \bibinfo{year}{2016}\natexlab{}.
\newblock \showarticletitle{Mining Duplicate Questions in Stack Overflow}. In
  \bibinfo{booktitle}{\emph{MSR}}.
\newblock


\bibitem[\protect\citeauthoryear{Awawdeh and Anderson}{Awawdeh and
  Anderson}{2010}]%
        {Awawdeh2010}
\bibfield{author}{\bibinfo{person}{R. Awawdeh} {and} \bibinfo{person}{T.
  Anderson}.} \bibinfo{year}{2010}\natexlab{}.
\newblock \showarticletitle{Improving Search in Tag-Based Systems with
  Automatically Extracted Keywords}. In \bibinfo{booktitle}{\emph{Knowledge
  Science, Engineering and Management}}.
\newblock


\bibitem[\protect\citeauthoryear{Bagheri and Ensan}{Bagheri and Ensan}{2016}]%
        {bagheri2016}
\bibfield{author}{\bibinfo{person}{E. Bagheri} {and} \bibinfo{person}{F.
  Ensan}.} \bibinfo{year}{2016}\natexlab{}.
\newblock \showarticletitle{Semantic Tagging and Linking of Software
  Engineering Social Content}.
\newblock \bibinfo{journal}{\emph{Automated Software Engineering}}
  (\bibinfo{year}{2016}).
\newblock


\bibitem[\protect\citeauthoryear{Barab{\'a}si and Albert}{Barab{\'a}si and
  Albert}{1999}]%
        {Barabasi1999}
\bibfield{author}{\bibinfo{person}{A.-L. Barab{\'a}si} {and}
  \bibinfo{person}{R. Albert}.} \bibinfo{year}{1999}\natexlab{}.
\newblock \showarticletitle{Emergence of Scaling in Random Networks}.
\newblock \bibinfo{journal}{\emph{Science}} (\bibinfo{year}{1999}).
\newblock


\bibitem[\protect\citeauthoryear{Cantador, Konstas, and Jose}{Cantador
  et~al\mbox{.}}{2011}]%
        {CANTADOR2011}
\bibfield{author}{\bibinfo{person}{I. Cantador}, \bibinfo{person}{I. Konstas},
  {and} \bibinfo{person}{J. Jose}.} \bibinfo{year}{2011}\natexlab{}.
\newblock \showarticletitle{Categorising Social Tags to Improve
  Folksonomy-based Recommendations}.
\newblock \bibinfo{journal}{\emph{Journal of Web Semantics}}
  (\bibinfo{year}{2011}).
\newblock


\bibitem[\protect\citeauthoryear{Cattuto, Benz, Hotho, and Stumme}{Cattuto
  et~al\mbox{.}}{2008}]%
        {cattuto2008}
\bibfield{author}{\bibinfo{person}{C. Cattuto}, \bibinfo{person}{D. Benz},
  \bibinfo{person}{A. Hotho}, {and} \bibinfo{person}{G. Stumme}.}
  \bibinfo{year}{2008}\natexlab{}.
\newblock \showarticletitle{Semantic Grounding of Tag Relatedness in Social
  Bookmarking Systems}. In \bibinfo{booktitle}{\emph{ISWC}}.
\newblock


\bibitem[\protect\citeauthoryear{{Chen}, {Alsakran}, {Barlowe}, {Yang}, and
  {Zhao}}{{Chen} et~al\mbox{.}}{2011}]%
        {Chen2011}
\bibfield{author}{\bibinfo{person}{Y. {Chen}}, \bibinfo{person}{J. {Alsakran}},
  \bibinfo{person}{S. {Barlowe}}, \bibinfo{person}{J. {Yang}}, {and}
  \bibinfo{person}{Y. {Zhao}}.} \bibinfo{year}{2011}\natexlab{}.
\newblock \showarticletitle{Supporting effective common ground construction in
  Asynchronous Collaborative Visual Analytics}. In
  \bibinfo{booktitle}{\emph{VAST}}.
\newblock


\bibitem[\protect\citeauthoryear{Chi}{Chi}{2008}]%
        {chi2008information}
\bibfield{author}{\bibinfo{person}{E. Chi}.} \bibinfo{year}{2008}\natexlab{}.
\newblock \showarticletitle{Information Seeking can be Social}.
\newblock \bibinfo{journal}{\emph{Information Seeking Support Systems}}
  (\bibinfo{year}{2008}).
\newblock


\bibitem[\protect\citeauthoryear{Chi and Mytkowicz}{Chi and Mytkowicz}{2008}]%
        {chi2008understanding}
\bibfield{author}{\bibinfo{person}{E. Chi} {and} \bibinfo{person}{T.
  Mytkowicz}.} \bibinfo{year}{2008}\natexlab{}.
\newblock \showarticletitle{Understanding the Efficiency of Social Tagging
  Systems using Information Theory}. In \bibinfo{booktitle}{\emph{ICWSM}}.
\newblock


\bibitem[\protect\citeauthoryear{de~Meo, Ferrara, Abel, Aroyo, and
  Houben}{de~Meo et~al\mbox{.}}{2014}]%
        {Meo2014}
\bibfield{author}{\bibinfo{person}{P. de Meo}, \bibinfo{person}{E. Ferrara},
  \bibinfo{person}{F. Abel}, \bibinfo{person}{L. Aroyo}, {and}
  \bibinfo{person}{G.-J. Houben}.} \bibinfo{year}{2014}\natexlab{}.
\newblock \showarticletitle{Analyzing User Behavior across Social Sharing
  Environments}.
\newblock \bibinfo{journal}{\emph{ACM TIST}} (\bibinfo{year}{2014}).
\newblock


\bibitem[\protect\citeauthoryear{Dimitrov, Helic, and Strohmaier}{Dimitrov
  et~al\mbox{.}}{2018}]%
        {dimitrov2018tag}
\bibfield{author}{\bibinfo{person}{D. Dimitrov}, \bibinfo{person}{D. Helic},
  {and} \bibinfo{person}{M. Strohmaier}.} \bibinfo{year}{2018}\natexlab{}.
\newblock \showarticletitle{Tag-based Navigation and Visualization}.
\newblock In \bibinfo{booktitle}{\emph{Social Information Access}}.
\newblock


\bibitem[\protect\citeauthoryear{Enrich, Braunhofer, and Ricci}{Enrich
  et~al\mbox{.}}{2013}]%
        {enrich2013}
\bibfield{author}{\bibinfo{person}{M. Enrich}, \bibinfo{person}{M. Braunhofer},
  {and} \bibinfo{person}{F. Ricci}.} \bibinfo{year}{2013}\natexlab{}.
\newblock \showarticletitle{Cold-start Management with Cross-Domain
  Collaborative Filtering and Tags}. In \bibinfo{booktitle}{\emph{ECWT}}.
\newblock


\bibitem[\protect\citeauthoryear{Furnas, Fake, Ahn, Schachter, Golder, Fox,
  Davis, Marlow, and Naaman}{Furnas et~al\mbox{.}}{2006}]%
        {furnas2006tagging}
\bibfield{author}{\bibinfo{person}{G.W. Furnas}, \bibinfo{person}{C. Fake},
  \bibinfo{person}{L. Ahn}, \bibinfo{person}{J. Schachter}, \bibinfo{person}{S.
  Golder}, \bibinfo{person}{K. Fox}, \bibinfo{person}{M. Davis},
  \bibinfo{person}{C. Marlow}, {and} \bibinfo{person}{M. Naaman}.}
  \bibinfo{year}{2006}\natexlab{}.
\newblock \showarticletitle{Why Do Tagging Systems Work?}. In
  \bibinfo{booktitle}{\emph{CHI}}.
\newblock


\bibitem[\protect\citeauthoryear{Gligori{\'c}, Anderson, and West}{Gligori{\'c}
  et~al\mbox{.}}{2018}]%
        {gligoric2018constraints}
\bibfield{author}{\bibinfo{person}{K. Gligori{\'c}}, \bibinfo{person}{A.
  Anderson}, {and} \bibinfo{person}{R. West}.} \bibinfo{year}{2018}\natexlab{}.
\newblock \showarticletitle{How Constraints Affect Content: The Case of
  Twitter's Switch from 140 to 280 Characters}. In
  \bibinfo{booktitle}{\emph{ICWSM}}.
\newblock


\bibitem[\protect\citeauthoryear{Golder and Huberman}{Golder and
  Huberman}{2006}]%
        {Golder2006}
\bibfield{author}{\bibinfo{person}{S.A. Golder} {and} \bibinfo{person}{B.A.
  Huberman}.} \bibinfo{year}{2006}\natexlab{}.
\newblock \showarticletitle{Usage patterns of collaborative tagging systems}.
\newblock \bibinfo{journal}{\emph{Journal of Information Science}}
  (\bibinfo{year}{2006}).
\newblock


\bibitem[\protect\citeauthoryear{Heaps}{Heaps}{1978}]%
        {heaps1978}
\bibfield{author}{\bibinfo{person}{H.~S. Heaps}.}
  \bibinfo{year}{1978}\natexlab{}.
\newblock \bibinfo{booktitle}{\emph{Information Retrieval: Computational and
  Theoretical Aspects}}.
\newblock \bibinfo{publisher}{Academic Press, Inc.}
\newblock


\bibitem[\protect\citeauthoryear{Helic and Strohmaier}{Helic and
  Strohmaier}{2011}]%
        {helic2011}
\bibfield{author}{\bibinfo{person}{D. Helic} {and} \bibinfo{person}{M.
  Strohmaier}.} \bibinfo{year}{2011}\natexlab{}.
\newblock \showarticletitle{Building Directories for Social Tagging Systems}.
  In \bibinfo{booktitle}{\emph{CIKM}}.
\newblock


\bibitem[\protect\citeauthoryear{Helic, Trattner, Strohmaier, and
  Andrews}{Helic et~al\mbox{.}}{2010}]%
        {helic2010navigability}
\bibfield{author}{\bibinfo{person}{D. Helic}, \bibinfo{person}{C. Trattner},
  \bibinfo{person}{M. Strohmaier}, {and} \bibinfo{person}{K. Andrews}.}
  \bibinfo{year}{2010}\natexlab{}.
\newblock \showarticletitle{On the Navigability of Social Tagging Systems}. In
  \bibinfo{booktitle}{\emph{ICSC}}.
\newblock


\bibitem[\protect\citeauthoryear{Heymann, Paepcke, and Garcia-Molina}{Heymann
  et~al\mbox{.}}{2010}]%
        {Heymann2010}
\bibfield{author}{\bibinfo{person}{P. Heymann}, \bibinfo{person}{A. Paepcke},
  {and} \bibinfo{person}{H. Garcia-Molina}.} \bibinfo{year}{2010}\natexlab{}.
\newblock \showarticletitle{Tagging Human Knowledge}. In
  \bibinfo{booktitle}{\emph{WSDM}}.
\newblock


\bibitem[\protect\citeauthoryear{Heymann, Ramage, and Garcia-Molina}{Heymann
  et~al\mbox{.}}{2008}]%
        {heymann2008social}
\bibfield{author}{\bibinfo{person}{P. Heymann}, \bibinfo{person}{D. Ramage},
  {and} \bibinfo{person}{H. Garcia-Molina}.} \bibinfo{year}{2008}\natexlab{}.
\newblock \showarticletitle{Social Tag Prediction}. In
  \bibinfo{booktitle}{\emph{SIGIR}}.
\newblock


\bibitem[\protect\citeauthoryear{Hotho, J{\"a}schke, Schmitz, and Stumme}{Hotho
  et~al\mbox{.}}{2006}]%
        {hotho2006}
\bibfield{author}{\bibinfo{person}{A. Hotho}, \bibinfo{person}{R. J{\"a}schke},
  \bibinfo{person}{C. Schmitz}, {and} \bibinfo{person}{G. Stumme}.}
  \bibinfo{year}{2006}\natexlab{}.
\newblock \showarticletitle{Information Retrieval in Folksonomies: Search and
  Ranking}. In \bibinfo{booktitle}{\emph{The Semantic Web: Research and
  Applications}}.
\newblock


\bibitem[\protect\citeauthoryear{Jabeen, Khusro, Majid, and Rauf}{Jabeen
  et~al\mbox{.}}{2014}]%
        {jabeen2014}
\bibfield{author}{\bibinfo{person}{F. Jabeen}, \bibinfo{person}{S. Khusro},
  \bibinfo{person}{A. Majid}, {and} \bibinfo{person}{A. Rauf}.}
  \bibinfo{year}{2014}\natexlab{}.
\newblock \showarticletitle{Semantics Discovery in Social Tagging Systems: A
  Review}.
\newblock \bibinfo{journal}{\emph{Multimedia Tools \& Applications}}
  (\bibinfo{year}{2014}).
\newblock


\bibitem[\protect\citeauthoryear{Kammerer, Nairn, Pirolli, and Chi}{Kammerer
  et~al\mbox{.}}{2009}]%
        {kammerer2009signpost}
\bibfield{author}{\bibinfo{person}{Y. Kammerer}, \bibinfo{person}{R. Nairn},
  \bibinfo{person}{P. Pirolli}, {and} \bibinfo{person}{E. Chi}.}
  \bibinfo{year}{2009}\natexlab{}.
\newblock \showarticletitle{Signpost from the Masses: Learning Effects in an
  Exploratory Social Tag Search Browser}. In \bibinfo{booktitle}{\emph{CHI}}.
\newblock


\bibitem[\protect\citeauthoryear{Kla{\v{s}}nja-Mili{\'c}evi{\'c}, Ivanovi{\'c},
  Vesin, and Budimac}{Kla{\v{s}}nja-Mili{\'c}evi{\'c} et~al\mbox{.}}{2018}]%
        {klavsnja2018}
\bibfield{author}{\bibinfo{person}{A. Kla{\v{s}}nja-Mili{\'c}evi{\'c}},
  \bibinfo{person}{M. Ivanovi{\'c}}, \bibinfo{person}{B. Vesin}, {and}
  \bibinfo{person}{Z. Budimac}.} \bibinfo{year}{2018}\natexlab{}.
\newblock \showarticletitle{Enhancing E-Learning Systems with Personalized
  Recommendation Based On Collaborative Tagging Techniques}.
\newblock \bibinfo{journal}{\emph{Applied Intelligence}}
  (\bibinfo{year}{2018}).
\newblock


\bibitem[\protect\citeauthoryear{Kleinberg, Kumar, Raghavan, Rajagopalan, and
  Tomkins}{Kleinberg et~al\mbox{.}}{1999}]%
        {kleinberg1999}
\bibfield{author}{\bibinfo{person}{J.M. Kleinberg}, \bibinfo{person}{R. Kumar},
  \bibinfo{person}{P. Raghavan}, \bibinfo{person}{S. Rajagopalan}, {and}
  \bibinfo{person}{A.S. Tomkins}.} \bibinfo{year}{1999}\natexlab{}.
\newblock \showarticletitle{The Web as a Graph: Measurements, Models, and
  Methods}. In \bibinfo{booktitle}{\emph{Computing and Combinatorics}}.
\newblock


\bibitem[\protect\citeauthoryear{K\"{o}rner, Benz, Hotho, Strohmaier, and
  Stumme}{K\"{o}rner et~al\mbox{.}}{2010}]%
        {koerner2010}
\bibfield{author}{\bibinfo{person}{C. K\"{o}rner}, \bibinfo{person}{D. Benz},
  \bibinfo{person}{A. Hotho}, \bibinfo{person}{M. Strohmaier}, {and}
  \bibinfo{person}{G. Stumme}.} \bibinfo{year}{2010}\natexlab{}.
\newblock \showarticletitle{Stop Thinking, Start Tagging: Tag Semantics Emerge
  from Collaborative Verbosity}. In \bibinfo{booktitle}{\emph{WWW}}.
\newblock


\bibitem[\protect\citeauthoryear{Kowald, Pujari, and Lex}{Kowald
  et~al\mbox{.}}{2017}]%
        {kowald2017temporal}
\bibfield{author}{\bibinfo{person}{D. Kowald}, \bibinfo{person}{S.C. Pujari},
  {and} \bibinfo{person}{E. Lex}.} \bibinfo{year}{2017}\natexlab{}.
\newblock \showarticletitle{Temporal Effects on Hashtag Reuse in Twitter: A
  Cognitive-Inspired Hashtag Recommendation Approach}. In
  \bibinfo{booktitle}{\emph{WWW}}.
\newblock


\bibitem[\protect\citeauthoryear{Krestel, Fankhauser, and Nejdl}{Krestel
  et~al\mbox{.}}{2009}]%
        {krestel2009latent}
\bibfield{author}{\bibinfo{person}{R. Krestel}, \bibinfo{person}{P.
  Fankhauser}, {and} \bibinfo{person}{W. Nejdl}.}
  \bibinfo{year}{2009}\natexlab{}.
\newblock \showarticletitle{Latent Dirichlet Allocation for Tag
  Recommendation}. In \bibinfo{booktitle}{\emph{RecSys}}.
\newblock


\bibitem[\protect\citeauthoryear{Ley and Seitlinger}{Ley and
  Seitlinger}{2015}]%
        {LEY2015}
\bibfield{author}{\bibinfo{person}{T. Ley} {and} \bibinfo{person}{P.
  Seitlinger}.} \bibinfo{year}{2015}\natexlab{}.
\newblock \showarticletitle{Dynamics of Human Categorization in a Collaborative
  Tagging System: How Social Processes of Semantic Stabilization Shape
  Individual Sensemaking}.
\newblock \bibinfo{journal}{\emph{Computers in Human Behavior}}
  (\bibinfo{year}{2015}).
\newblock


\bibitem[\protect\citeauthoryear{Lin, Trattner, Brusilovsky, and He}{Lin
  et~al\mbox{.}}{2015}]%
        {lin2015}
\bibfield{author}{\bibinfo{person}{Y.-L. Lin}, \bibinfo{person}{C. Trattner},
  \bibinfo{person}{P. Brusilovsky}, {and} \bibinfo{person}{D. He}.}
  \bibinfo{year}{2015}\natexlab{}.
\newblock \showarticletitle{The Impact of Image Descriptions on User Tagging
  Behavior: A Study of the Nature and Functionality of Crowdsourced Tags}.
\newblock \bibinfo{journal}{\emph{Journal of the Association for Information
  Science and Technology}} (\bibinfo{year}{2015}).
\newblock


\bibitem[\protect\citeauthoryear{Mamykina, Miller, Grevet, Medynskiy, Terry,
  Mynatt, and Davidson}{Mamykina et~al\mbox{.}}{2011}]%
        {Mamykina2011}
\bibfield{author}{\bibinfo{person}{L. Mamykina}, \bibinfo{person}{A.D. Miller},
  \bibinfo{person}{C. Grevet}, \bibinfo{person}{Y. Medynskiy},
  \bibinfo{person}{M.A. Terry}, \bibinfo{person}{E.D. Mynatt}, {and}
  \bibinfo{person}{P.R. Davidson}.} \bibinfo{year}{2011}\natexlab{}.
\newblock \showarticletitle{Examining the Impact of Collaborative Tagging on
  Sensemaking in Nutrition Management}. In \bibinfo{booktitle}{\emph{CHI}}.
\newblock


\bibitem[\protect\citeauthoryear{Markines, Cattuto, Menczer, Benz, Hotho, and
  Stumme}{Markines et~al\mbox{.}}{2009}]%
        {markines2009}
\bibfield{author}{\bibinfo{person}{B. Markines}, \bibinfo{person}{C. Cattuto},
  \bibinfo{person}{F. Menczer}, \bibinfo{person}{D. Benz}, \bibinfo{person}{A.
  Hotho}, {and} \bibinfo{person}{G. Stumme}.} \bibinfo{year}{2009}\natexlab{}.
\newblock \showarticletitle{Evaluating Similarity Measures for Emergent
  Semantics of Social Tagging}. In \bibinfo{booktitle}{\emph{WWW}}.
\newblock


\bibitem[\protect\citeauthoryear{Mendes, Ringrose, and Keller}{Mendes
  et~al\mbox{.}}{2018}]%
        {mendes2018metoo}
\bibfield{author}{\bibinfo{person}{K. Mendes}, \bibinfo{person}{J. Ringrose},
  {and} \bibinfo{person}{J. Keller}.} \bibinfo{year}{2018}\natexlab{}.
\newblock \showarticletitle{\#MeToo and the Promise and Pitfalls of Challenging
  Rape Culture Through Digital Feminist Activism}.
\newblock \bibinfo{journal}{\emph{European Journal of Women's Studies}}
  (\bibinfo{year}{2018}).
\newblock


\bibitem[\protect\citeauthoryear{Miotto and Weng}{Miotto and Weng}{2013}]%
        {MIOTTO2013}
\bibfield{author}{\bibinfo{person}{R. Miotto} {and} \bibinfo{person}{C. Weng}.}
  \bibinfo{year}{2013}\natexlab{}.
\newblock \showarticletitle{Unsupervised Mining of Frequent Tags for Clinical
  Eligibility Text Indexing}.
\newblock \bibinfo{journal}{\emph{Journal of Biomedical Informatics}}
  (\bibinfo{year}{2013}).
\newblock


\bibitem[\protect\citeauthoryear{Newman}{Newman}{2005}]%
        {newman2005}
\bibfield{author}{\bibinfo{person}{M.E.J. Newman}.}
  \bibinfo{year}{2005}\natexlab{}.
\newblock \showarticletitle{Power laws, Pareto distributions and Zipf's law}.
\newblock \bibinfo{journal}{\emph{Contemporary Physics}}
  (\bibinfo{year}{2005}).
\newblock


\bibitem[\protect\citeauthoryear{Olteanu, Weber, and Gatica-Perez}{Olteanu
  et~al\mbox{.}}{2016}]%
        {olteanu2016characterizing}
\bibfield{author}{\bibinfo{person}{A. Olteanu}, \bibinfo{person}{I. Weber},
  {and} \bibinfo{person}{D. Gatica-Perez}.} \bibinfo{year}{2016}\natexlab{}.
\newblock \showarticletitle{Characterizing the Demographics Behind the
  \#BlackLivesMatter Movement}. In \bibinfo{booktitle}{\emph{AAAI Spring
  Symposium Series}}.
\newblock


\bibitem[\protect\citeauthoryear{Schmitz, Hotho, J{\"a}schke, and
  Stumme}{Schmitz et~al\mbox{.}}{2006}]%
        {schmitz2006}
\bibfield{author}{\bibinfo{person}{C. Schmitz}, \bibinfo{person}{A. Hotho},
  \bibinfo{person}{R. J{\"a}schke}, {and} \bibinfo{person}{G. Stumme}.}
  \bibinfo{year}{2006}\natexlab{}.
\newblock \showarticletitle{Mining Association Rules in Folksonomies}. In
  \bibinfo{booktitle}{\emph{Data Science and Classification}}.
\newblock


\bibitem[\protect\citeauthoryear{Simkin and Roychowdhury}{Simkin and
  Roychowdhury}{2011}]%
        {SIMKIN20111}
\bibfield{author}{\bibinfo{person}{M.V. Simkin} {and} \bibinfo{person}{V.P.
  Roychowdhury}.} \bibinfo{year}{2011}\natexlab{}.
\newblock \showarticletitle{Re-inventing Willis}.
\newblock \bibinfo{journal}{\emph{Physics Reports}} (\bibinfo{year}{2011}).
\newblock


\bibitem[\protect\citeauthoryear{Trattner, Kowald, Seitlinger, Ley, and
  Kopeinik}{Trattner et~al\mbox{.}}{2016}]%
        {trattner2016}
\bibfield{author}{\bibinfo{person}{C. Trattner}, \bibinfo{person}{D. Kowald},
  \bibinfo{person}{P. Seitlinger}, \bibinfo{person}{T. Ley}, {and}
  \bibinfo{person}{S. Kopeinik}.} \bibinfo{year}{2016}\natexlab{}.
\newblock \showarticletitle{Modeling Activation Processes in Human Memory to
  Predict the Use of Tags in Social Bookmarking Systems}.
\newblock \bibinfo{journal}{\emph{The Journal of Web Science}}
  (\bibinfo{year}{2016}).
\newblock


\bibitem[\protect\citeauthoryear{Tria, Loreto, Servedio, and Strogatz}{Tria
  et~al\mbox{.}}{2013}]%
        {tria2013}
\bibfield{author}{\bibinfo{person}{F. Tria}, \bibinfo{person}{V. Loreto},
  \bibinfo{person}{V.D.P. Servedio}, {and} \bibinfo{person}{S.H. Strogatz}.}
  \bibinfo{year}{2013}\natexlab{}.
\newblock \showarticletitle{The dynamics of correlated novelties.}
\newblock \bibinfo{journal}{\emph{CoRR}} (\bibinfo{year}{2013}).
\newblock


\bibitem[\protect\citeauthoryear{Willett, Heer, Hellerstein, and
  Agrawala}{Willett et~al\mbox{.}}{2011}]%
        {Willet2011}
\bibfield{author}{\bibinfo{person}{W. Willett}, \bibinfo{person}{J. Heer},
  \bibinfo{person}{J. Hellerstein}, {and} \bibinfo{person}{M. Agrawala}.}
  \bibinfo{year}{2011}\natexlab{}.
\newblock \showarticletitle{CommentSpace: Structured Support for Collaborative
  Visual Analysis}. In \bibinfo{booktitle}{\emph{CHI}}.
\newblock


\bibitem[\protect\citeauthoryear{Xie, Li, Mao, Li, Cai, and Rao}{Xie
  et~al\mbox{.}}{2014}]%
        {XIE2014}
\bibfield{author}{\bibinfo{person}{H. Xie}, \bibinfo{person}{Q. Li},
  \bibinfo{person}{X. Mao}, \bibinfo{person}{X. Li}, \bibinfo{person}{Y. Cai},
  {and} \bibinfo{person}{Y. Rao}.} \bibinfo{year}{2014}\natexlab{}.
\newblock \showarticletitle{Community-aware User Profile Enrichment in
  Folksonomy}.
\newblock \bibinfo{journal}{\emph{Neural Networks}} (\bibinfo{year}{2014}).
\newblock


\bibitem[\protect\citeauthoryear{Yule}{Yule}{1925}]%
        {yule1925}
\bibfield{author}{\bibinfo{person}{G.U. Yule}.}
  \bibinfo{year}{1925}\natexlab{}.
\newblock \showarticletitle{A Mathematical Theory of Evolution, Based on the
  Conclusions of Dr. Willis}.
\newblock \bibinfo{journal}{\emph{Transactions of the Royal Society of London}}
  (\bibinfo{year}{1925}).
\newblock


\bibitem[\protect\citeauthoryear{Zhang}{Zhang}{2019}]%
        {zhang2019language}
\bibfield{author}{\bibinfo{person}{Y. Zhang}.} \bibinfo{year}{2019}\natexlab{}.
\newblock \showarticletitle{Language in Our Time: An Empirical Analysis of
  Hashtags}. In \bibinfo{booktitle}{\emph{WWW}}.
\newblock


\bibitem[\protect\citeauthoryear{Zhang, Zhou, and Zhang}{Zhang
  et~al\mbox{.}}{2011}]%
        {zhang2011}
\bibfield{author}{\bibinfo{person}{Z.-K. Zhang}, \bibinfo{person}{T. Zhou},
  {and} \bibinfo{person}{Y.-C. Zhang}.} \bibinfo{year}{2011}\natexlab{}.
\newblock \showarticletitle{Tag-aware Recommender Systems: A State-of-the-art
  Survey}.
\newblock \bibinfo{journal}{\emph{Journal of Computer Science and Technology}}
  (\bibinfo{year}{2011}).
\newblock


\bibitem[\protect\citeauthoryear{Zubiaga, K\"{o}rner, and Strohmaier}{Zubiaga
  et~al\mbox{.}}{2011}]%
        {Zubiaga2011}
\bibfield{author}{\bibinfo{person}{A. Zubiaga}, \bibinfo{person}{C.
  K\"{o}rner}, {and} \bibinfo{person}{M. Strohmaier}.}
  \bibinfo{year}{2011}\natexlab{}.
\newblock \showarticletitle{Tags vs Shelves: From Social Tagging to Social
  Classification}. In \bibinfo{booktitle}{\emph{HT}}.
\newblock


\bibitem[\protect\citeauthoryear{Zuo, Zeng, Gong, and Jiao}{Zuo
  et~al\mbox{.}}{2016}]%
        {zuo2016}
\bibfield{author}{\bibinfo{person}{Yi Zuo}, \bibinfo{person}{J. Zeng},
  \bibinfo{person}{M. Gong}, {and} \bibinfo{person}{L. Jiao}.}
  \bibinfo{year}{2016}\natexlab{}.
\newblock \showarticletitle{Tag-aware Recommender Systems Based On Deep Neural
  Networks}.
\newblock \bibinfo{journal}{\emph{Neurocomputing}} (\bibinfo{year}{2016}).
\newblock


\end{thebibliography}

\end{document}